%% file: cr.tex
\documentclass{elsart}

\usepackage{graphicx}
\usepackage{amssymb}
\usepackage{amsmath}
\usepackage{algorithmic}

\include{macrosAPS}

\newcommand{\mach}{{\mathcal M}}
\newcommand{\oo}{{\rm o}}

\begin{document}
\begin{frontmatter}
\title{Glimm-Godunov's Method for Cosmic-ray-hydrodynamics}

\author{Francesco Miniati}
\address{Physics Department, Wolfgang-Pauli-Strasse 16, ETH-Z\"urich, CH-8093 Z\"urich}

\begin{abstract}
  A numerical method for integrating the equations describing a
  dynamically coupled system made of a fluid and cosmic-rays is
  developed.  In smooth flows the effect of CR pressure is accounted
  for by modification of the characteristic equations and the energy
  exchange between cosmic-rays and the fluid, due to diffusive
  processes in configuration and momentum space, is modeled with a
  flux conserving method.  Provided the shock acceleration efficiency
  as a function of the upstream conditions and shock Mach number, we
  show that the Riemann solver can be modified to take into account
  the cosmic-ray mediation without having to resolve the cosmic-ray
  induced substructure.  Shocks are advanced with Glimm's method which
  preserves their discontinuous character without any smearing, thus
  allowing to maintain self-consistency in the shock solutions. In
  smooth flows either Glimm's or a higher order Godunov's method can
  be applied, with the latter producing better results when
  approximations are introduced in the Riemann solver.

\end{abstract}
\end{frontmatter}

\section{Introduction} \label{intro.sec}

We wish to formulate a numerical method to solve a system of equations
characterizing a fluid that is dynamically coupled to suprathermal
particles through the exchange of momentum and energy. Such conditions
occur commonly in astrophysical plasmas. The fluid system is an
ordinary nonrelativistic gas described by the following modified
equations of hydrodynamics :
\begin{eqnarray}
\label{rhoe:eq}
\frac{\partial \rho}{\partial t} + \frac{\partial}{\partial x_d} (\rho u_d) & = & 0, \\
 \label{mome:eq}
 \frac{\partial \rho u_i}{\partial t} + 
 \frac{\partial}{\partial x_d} [\rho u_i u_d +P_g\, \delta_{id}] 
 & = &  -\frac{\partial P_c}{\partial x_d}\,\delta_{id},
 \\ \label{enee:eq}
\frac{\partial \rho e_g}{\partial t} + \frac{\partial}{\partial x_d} [(\rho e_g+P_g) u_d]   & =  &  
-u_d \frac{\partial P_c}{\partial x_d} - \Sigma ,
\end{eqnarray}
where $(\rho, u, P_g, e_g)$ indicate the gas density, velocity,
pressure and specific energy respectively; $i,d$ index the spatial
components and summation over repeated indexes is assumed;
$\delta_{id} $ is the Kronecker's delta.  The gas total specific
energy is, $ e_g = u^2/2 + e_{th}$, and a $\gamma$-law equation of
state is assumed so that the gas specific internal energy is related
to the gas pressure through $e_{th} = P_g/\rho (\gamma_g-1)$.  The
inhomogeneous terms proportional to $\partial P_c/\partial x$ on the
right hand side of Eq.~(\ref{mome:eq})-(\ref{enee:eq}) account for the
effects of the suprathermal pressure.  $\Sigma$ is a source term
describing the transfer of energy between the fluid and the
suprathermal component.  This may be due to, e.g., particle
acceleration processes at the expenses of the fluid energy or,
conversely, energy losses from the suprathermal particles that end up
heating the fluid.

As for the suprathermal component we consider cosmic-ray (heretofore
CR) particles described by a distribution function, $f({\bf x},p,t)$,
which depends upon a spatial, a momentum and a temporal coordinate.
In what follows $p$ is in units of `$m_c c$', with $m_c$ the CR particle
mass, and the normalization of $f$ is such that the number density of
particles with momentum between $p$ and $p+dp$ is $dn_c=4\pi\, p^2\,
f\,dp$. In addition, $f$ is assumed to be isotropic in momentum space
and evolves according to the following diffusion-convection
equation~\cite{skill75a}
\begin{equation}  \label{dce:eq}
\frac{\partial f}{\partial t}  + u_d \frac{\partial f}{\partial x_d}-
\frac{\partial}{\partial x_d}\left( \kappa \frac{\partial f}{\partial x_d} \right)
= \frac{1}{3}\, \frac{\partial u_d}{\partial x_d}\, p\,\frac{\partial f}{\partial p} 
+  \frac{1}{p^2} \frac{\partial}{\partial p} \left[ p^2 \left(b_\ell\,f +
D_p \frac{\partial f}{\partial p} \right) \right]  .
\end{equation}
The second and third term on the left hand side of the above equation
represent, respectively, spatial advection and diffusion with a
coefficient, $\kappa ({\bf x},p)$. The first term on the right hand
side accounts for adiabatic effects and, $b_\ell(p)\equiv -(dp/dt)_{loss}$, 
describes the particle momentum change due to energy losses associated with
mechanical and radiative processes. In addition, $D_p (p)$ is the
diffusion coefficient in momentum space.  The CR pressure in
Eq.~(\ref{mome:eq}) and~(\ref{enee:eq}) is then defined through the
distribution function, $f$, as
\begin{equation} \label{pc:eq}
  P_c ({\bf x})=   \frac{4\pi}{3}\,m_cc^2 \int_{p_{min}}^{p_{max}}
  p^4 \, f({\bf x},p)\, (p^2 + 1)^{-\frac{1}{2}}\, dp ,
\end{equation}
where $p_{min},p_{max}$ are the minimum and maximum CR momenta,
respectively. More specifically, the momentum $p_{min}$ marks
(somewhat loosely) the transition between the thermal and nonthermal 
components and $p_{max}$ is the maximum momentum the particles
can achieve and still be confined inside the system.

The CR energy and adiabatic index are given by
\begin{eqnarray} \label{ec:eq}
E_c  &=& 4\pi\,m_cc^2\, \int_{p_{min}}^{p_{max}} p^2\, 
f({\bf x},p)\, [(p^2 + 1)^{\frac{1}{2}}-1] \,dp, \\ \label{gc:eq}
\gamma_c & =&  1+ \frac{P_c}{E_c}.
\end{eqnarray}
The evolution of the CR energy is obtained from Eq.~(\ref{dce:eq})
and reads
\begin{eqnarray} \nonumber
\frac{\partial E_c}{\partial t} &=& 
-\frac{\partial}{\partial x_d} \left(E_c u_d \right) 
+\frac{\partial}{\partial x_d} 
\left(\langle\kappa\rangle\frac{\partial E_c}{\partial x_d} \right) 
-P_c \frac{\partial u_d}{\partial x_d} \\ \nonumber
&& - 4\pi m_cc^2 \int_{p_{min}}^{p_{max}}
\left(b_{\ell} \,f + D_{p }\frac{\partial f}{\partial p} \right)
\frac{p^3}{(p^2 + 1)^{\frac{1}{2}}} \, dp 
\\ \label{ec_ev:eq}
&&+ 4\pi m_c c^2 \left\{p^2 f
\left(\frac{1}{3}\frac{\partial u_d}{\partial x_d} + b_{\ell} + D_{p }\frac{\partial \ln f}{\partial p}\right)
\left[(p^2+1)^{\frac{1}{2}}-1\right]\right\}_{p_{min}}^{p_{max}}.
\end{eqnarray}
The first line of Eq.~(\ref{ec_ev:eq}) refers to the effects of
advection, diffusion (with an energy averaged diffusion coefficient
$\langle\kappa\rangle$) and adiabatic compression, and the second
line to energy losses/gains introduced above.  The surface terms on the
third line describe changes in the CR energy due to the flux of
particles across the low and high boundaries in momentum space. The
first of these two surface terms is typically negligible whereas the
second is important in case of efficient shock acceleration and can
cause significant energy losses in the system.
Finally the source term in Eq.~(\ref{enee:eq}) is related to the
change in the CR distribution function due to flux in momentum space
as
\begin{eqnarray} \nonumber
\Sigma({\bf x}) = 
&& - 4\pi m_cc^2 \int_{p_{min}}^{p_{max}}
\left(b_{m\ell} \,f + D_{p }\frac{\partial f}{\partial p} \right)
\frac{p^3}{(p^2 + 1)^{\frac{1}{2}}} \, dp
\\  \label{sigmaeq:eq}
&& - 4\pi m_c c^2 p^2 f
\left(\frac{1}{3}\frac{\partial u_d}{\partial x_d} + b_{\ell} + D_{p }
\frac{\partial \ln f}{\partial p} \right)
\left.\left[(p^2+1)^{\frac{1}{2}}-1\right] \right|_{p=p_{min}}
\end{eqnarray}
where $b_{m\ell}(p)$ now includes mechanical losses only (i.e. radiative
losses are excluded).

The spatial diffusion coefficient introduces a physical scale
characterizing the particles mean free path due to diffusion, i.e.
$\lambda_{mfp} \sim \kappa(p)/v(p)$, where $v(p)\sim c$, the velocity
of a particle of momentum $p$, is of order the speed of light.  In the
following we shall distinguish two different regimes of application of
the equations (\ref{rhoe:eq})-(\ref{dce:eq}), namely smooth flows and
shock waves.  The reason for doing so is that for astrophysical
systems, $\lambda_{mfp}\ll \lambda_{system}$, and the entire dynamic
range of scales cannot be resolved with currently available computers.
However, while on the scales that can be resolved by simulations
diffusion in smooth flows can be safely assumed to become either slow
or negligible, this is not the case for shocks.

In smooth flows (and on large enough scales, $\lambda \gg
\lambda_{mfp}$) the presence of CRs enhances the propagation speed of
sound waves but simultaneously causes a damping of their amplitude due
to CR diffusion~\cite{parker65,ptuskin81}. In addition energy is
exchanged non adiabatically between the thermal and nonthermal
components according to Eq.~(\ref{sigmaeq:eq}). These effects arise
from diffusive processes and as long as the relevant transport
coefficients, $\kappa$ and $D_p$, are defined correctly, they can be
properly modeled numerically with schemes available in the literature.

Around shocks, however, the situation is more complicated because the
diffusion process gives rise to an efficient mechanism for
transferring energy from the flow to the particles.  This topic is
discussed in detail in several review articles~\cite{drury83,blei87}.
Here we emphasize two basic and related points relevant for the
present discussion. Firstly, the dissipation of energy into CRs
changes the value of the total pressure generated by the shock
dissipation mechanism due to the different thermodynamic properties of
gas and CRs.  In addition, as illustrated above (cf.
Eq.~[\ref{ec_ev:eq}]) the escape of high energy particles upstream of
the shock allows for energy to be removed from the system. This can
reduce the pressure support in the downstream region, allowing for
compression ratios higher than the hydrodynamic limit.  Finally, the
CR pressure gradient produced by the CR particles diffusing upstream
decelerates the flow approaching the shock front. As a result the
velocity structure is not a sharp transition anymore but is broadened
up to scales of order the diffusive scale length of the most energetic
CR particles. This is $\lambda_\kappa(p_{max})=
\kappa(p_{max})/u_{shock}$, where $u_{shock}$ is the shock speed.
This effect creates the so called shock {\it precursor} where the
upstream gas is adiabatically compressed before being shocked.  Thus,
even though in a numerical calculation the CRs do not diffuse out of
the resolution element during a timestep, shock acceleration can
modify significantly the shock jump conditions with respect to the
simple fluid case (see Section~\ref{rpcrh:se}).

There are, therefore, at least two different limits of interest for
solving the equations (\ref{rhoe:eq})-(\ref{dce:eq}) in the case of
shocks.  One which focuses on the study of the diffusive shock
acceleration process itself.  In this case one requires: ($i$) a
kinetic approach in which the evolution of the distribution function
in momentum space given by Eq.~(\ref{dce:eq}) is followed
accurately~\cite{acblpe84,malkov97}; ($ii$) enough spatial resolution
to properly resolve the full range of relevant scales that enter the
problem, from the thickness of the shock, $\ell_{shock}$, to the
diffusive scale length of the highest energy CR particles,
$\lambda_\kappa(p_{max})$. A number of codes, with different levels of
sophistication, employing various numerical methods and devoted to
this type of approach have been developed~(e.g.
\cite{elei84,ac87,kajo91,duffy92,beyeks94,giacalone97,kjlvs00,joka05}).
Sometime and to various extents they also include the processes that
regulate the diffusive properties of the medium (e.g. wave
amplification and damping), a key factor for the efficiency of the
acceleration mechanism. By necessity, they focus on a very narrow
region around the shock, of order $\lambda_\kappa(p_{max})$.
Complemented by analytic
studies~\cite{eichler84,mavo95,malkov97,malkov98}, among their
ultimate goals is to investigate, as a function of the upstream gas
conditions, $U^-$, and the shock Mach number, $\mach$, the downstream
CR distribution function, $f^+(p)$, and the efficiency of the shock
acceleration process, $\eta$. Here and in the following this quantity
is defined as the fraction of total pressure upstream of the shock
that is converted into downstream CR pressure, $P_c^+$:
\begin{equation} \label{eta:eq}
\eta(U^-,\mach) \equiv \frac{P_c^+}{P^-_g + P^-_c + \rho^-(u^-)^2} ,
\end{equation}
where $P_c^+$ is related to $f^+$ through Eq.~(\ref{pc:eq}).

There is then the opposite limit to the one described above, which
focuses on the dynamical effects of CRs in smooth flows and shocks,
for systems of size $\lambda_{system}\gg \lambda_\kappa(p_{max})$.  We
can refer to it as the {\it astrophysical} limit.  This approach is
more application
oriented~\cite{jre99,mrkj01,mjkr01,min02,min03,tregillis04,jubelgas06}
and does not aim at studying $f^+(p;U^-,\mach)$ or $\eta(U^-,\mach)$.
In this paper we attempt to design a numerical method that serves this
purpose, without having to resolve scales of order $\ell_{shock}$.  In
doing so, we seek to eliminate all but the essential information about
the CR distribution function in momentum space, so that a fluid-like
description is approached.  Information at the kinetic level must,
however, be preserved in two parts of the formulation: (a) when
computing shock solutions and (b) when computing the time-evolution of
the CR pressure. The first requirement is set because, as already
pointed out, a correct shock solution can only be obtained with a
fully kinetic approach~\cite{acblpe84,malkov97}. This means that the
only way to meaningfully include the effects of CRs acceleration in a
fluid-like approach is to assume that $f^+(p;U^-,\mach)$ and the shock
acceleration efficiency, $\eta(U^-,\mach)$, are provided independently
(e.g. from kinetic models) as part of the input.  When the
acceleration efficiency is high and the contribution from the highest
energy particles dominate the CR energy and pressure, the flux of CR
particles across $p_{max}$ must also be specified~\cite{mavo96}.  The
second requirement stems from the fact that $dP_c/dt$ is basically the
result of energy losses/gains and diffusion of the CR particles.
Since these processes are strongly momentum dependent, $dP_c/dt$ is
bound to be different for different shapes of the CR distribution
function and one ought to be able to account for this.  Note that the
two-fluid approximation alone, in which Eq.~(\ref{dce:eq}) is
integrated in momentum space to derive an equation for the time
evolution of $P_c$, is not sufficient for these two purposes.

In order to properly evolve $P_c$, we divide momentum space in a set
of $N_p$ ($\sim10$) coarse kinetic volumes or momentum {\it bins} and
integrate Eq.~(\ref{dce:eq}) within the boundaries of each of them.
This provides an equation for the evolution of the number density of
CRs within each bin and is a cost-effective way of following the
change of shape of the CR distribution function resulting from the
momentum dependent CR processes mentioned above.  This essentially
works because the CR distribution, $f$, is typically a smooth power
law with a slowly varying slope as a function of momentum.  This fact
provides a natural way for describing $f(p)$ within each
bin~\cite{min01,jre99}, which compensates for the coarseness of the
discretization of momentum space.

In addition, we show that once the shock acceleration efficiency,
$\eta$, is specified it is possible to account for the modifications
induced by the CRs on the hydrodynamic shock solution, even though the
structure of the shocks remains completely unresolved.  This is
achieved by solving a slightly more complicated Riemann problem, after
proper modification of the definition of the nonlinear waves that
appear in it.  This shock solution can still be thought of in terms of
a two-fluid model description~\cite{drvo81} but with the important
difference that, among the family of admissible
solutions~\cite{acblpe84}, we select the one demanded by the
(explicitly) adopted shock acceleration model.  This ensures a self
consistent description of the CR-hydrodynamic system.

In order for this to work, however, it is essential that the shock
discontinuity does not spread as a result of numerical dissipation.
This is because in general the dissipation of CR energy at a shock is
a nonlinear function of the full jump conditions. Therefore, if the
shock is artificially spread over a few zones the sum of the CR energy
generated at each numerical subshock will in general not be the same
as that predicted by the model for the full jump conditions.

A suitable hydrodynamic method for our purpose is the one originally
proposed by Glimm~\cite{glimm65}.  Introduced as part of a
constructive proof of existence of solution to nonlinear hyperbolic
conservation laws it was turned into an effective numerical scheme for
hydrodynamics by Chorin~\cite{chorin76,chorin77}.  Glimm's method
maintains unsmoothed all the sharp features that are present in the
flow. In particular, shocks remain unsmoothed jumps as they propagate
across the grid.  This allows us to maintain self-consistency in the
shock solution. The limitation in using Glimm's method is that at the
moment its multidimensional extension does not work properly at
shocks~\cite{colella82}.  Thus here we focus on a one dimensional
algorithm and leave its generalization to more than one dimension for
future work.

In smooth flows either Glimm's or Godunov's method can be applied.
Therefore it is possible to define a hybrid scheme where Glimm's
method is applied at shocks and Godunov's method in smooth parts of
the flow.  In either case, (in smooth flows) the effects due to the CR
pressure are included by proper modification of the characteristic
analysis.

Note that recently a method has been proposed
in~\cite{jubelgas06,pfrommer06} to include the dynamical effect of CR
pressure on the hydrodynamics.  That approach consists effectively of
a two fluid model in which the generation of CRs at shocks is treated
as an explicit source term, similar to the scheme in~\cite{min01}.
However, a hydro scheme that includes self-consistently the kinetic
effects on both the shock substructure and the CR pressure evolution,
in the sense mentioned in (a) and (b) above, is still lacking.

This paper is structured as follows.  In Section~\ref{dms:se} we
describe the discretization of momentum space and compute the fluxes
due to energy losses and diffusion in that dimension.  In
Section~\ref{rpcrh:se} we discuss the effects of CRs on the structure
of the flow and define a modified Riemann problem to include the
effects of CRs both in smooth flows and at shocks.  In
Section~\ref{glimm:se} and ~\ref{ggm:se} we describe the
implementations of Glimm's method and of a hybrid Glimm-Godunov's
method, respectively.  Tests follow in Section~\ref{tests:se} and
conclusions are presented in Section~\ref{concl:se}.

\section{Diffusion-Convection Equation} \label{dms:se}
In order to formulate a finite-kinetic-volume description of the
diffusion-convection equation~(\ref{dce:eq}), we divide momentum space
into $N_p$ logarithmically spaced bins, each with boundaries
$p_{j-\frac{1}{2}},~p_{j+\frac{1}{2}}$. The log-width of the bins,
$\Delta w_j \equiv \log(p_{j+\frac{1}{2}}/p_{j-\frac{1}{2}})$, is
taken as constant (although this is not necessary). We then follow the
evolution of the CR number density associated with each bin, namely
\begin{equation} \label{nbin:eq}
n_{p_j}  =  \int_{p_{j-\frac{1}{2}}}^{p_{j+\frac{1}{2}}} 4\pi\,p^2 \;f(p)\; dp.
\end{equation}
For a piecewise power-law distribution function we have
\begin{eqnarray} \label{ppl:eq}
f(p)& = & f_j(p)  = f_{0j} \; \left(\frac{p}{p_{j-1}}\right)^{-q_j}, \quad p_{j-\frac{1}{2}}\leq p < p_{j+\frac{1}{2}} \, , \\
\label{kave:eq}
n_{p_j} & = & 4\pi \;f_{0j}\;p_{j-\frac{1}{2}}^3\;
\frac{(p_{j+\frac{1}{2}}/p_{j-\frac{1}{2}})^{3-q_j}-1}{3-q_j} ,
\end{eqnarray}
where $f_{0j}$ and $q_j$ are the normalization and logarithmic slope
for $f$ in the $\jmath_{th}$ momentum bin.  Once the set, $\{n_{p_j};
0\le j <N_p\}$, is defined and the boundary conditions for the slopes
$q_{-1}$ and $q_{N_p}$ are provided, we can compute the set of slopes,
$\{q_j; 0\le j <N_p\}$, and normalizations, $\{f_{0j}; 0\le j <N_p\}$,
as follows~\cite{jre99}.  For each bin, $i$, we use
Eq.~(\ref{kave:eq}) with $j=i,i\pm 1$ and further assume: (a) that the
spectral curvature, $\partial q/\partial \ln p$, remains constant
across adjacent bins; (b) that $f(p)$, as given in Eq.~(\ref{ppl:eq}),
is continuous across the bins boundaries, $p_{i-\frac{1}{2}}$ and $p_{i+\frac{1}{2}}$.
This provides six equations for six variables that can be efficiently
solved with an iterative method. This procedure is applied for each
bin, and allows to reconstruct the set $\{f_j(p)\}$ from
$\{n_{p_j}\}$.  In the following we will use both $n_{p_j}$ and
$f_j(p)$ assuming that we can reconstruct the latter from the former
through the procedure just described\footnote{When spectral curvature
  is important, the following alternative approach first proposed
  in~\cite{min01} can be used instead: for each bin, in addition to
  the number density of CR particles, one also follows the energy
  density, $\epsilon_{p_j}$.  For each bin the definitions of
  $n_{p_j}$ and $\epsilon_{p_j}$ provide two equations which can be
  readily solved for the two unknowns $(f_{0j},q_j)$.  While the
  equation for $n_{p_j}$ is derived below (cf Eq.~\ref{dcei:eq}), the one
  for $\epsilon_{p_j}$ is obtained analogously by multiplying
  Eq.~(\ref{dce:eq}) by a factor, $4\pi\,p^2[(p^2+1)^{\frac{1}{2}}-1]$, and
  integrating it within the interval, $p_{j-\frac{1}{2}}-p_{j+\frac{1}{2}}$. See
  also~\cite{joka05} for the effectiveness of this method. Here we take
  the simpler approach, however, in which we follow $n_{p_j}$ only.
  This is because we wish to focus on the novelty of the method which
  is related to the fluid aspect of the solutions. In fact, the method
  for the evolution of $f(p)$ in phase space was extensively studied
  in~\cite{min01}.}.

The equation describing the evolution of, $n_{p_j}$, is obtained by
integrating Eq.~(\ref{dce:eq}), multiplied by a factor $4\pi\,p^2$,
within the interval, $p_{j-\frac{1}{2}}-p_{j+\frac{1}{2}}$. This leads to
\begin{eqnarray}
\label{dcei:eq} 
\frac{\partial n_{p_j}}{\partial t} & + & 
 {\bf \nabla }_x \cdot F_x = - \tilde{\bf \nabla}_p F_p + J_j ,  \\
\label{fxe:eq}
F_x  & = & {\bf u}\,n_{p_j} - {\bf \langle \kappa \rangle \nabla }n_{p_j}, \\
\label{fpe:eq}
F_p  & = & 4\pi p^2 f_j(p) \, \dot{p }
\end{eqnarray}
where $\tilde{\bf \nabla}_p$ is the undivided (one-dimensional)
gradient in momentum space and we have introduced
\begin{eqnarray} 
\langle \kappa \rangle _j & = &
 \int_{p_{j-\frac{1}{2}}}^{p_{j+\frac{1}{2}}} p^2 \kappa \nabla f_j(p)\; dp ~~/~ \int_{p_{j-\frac{1}{2}}}^{p_{j+\frac{1}{2}}} p^2 \nabla f_j(p)\; dp, \\
\dot{p} & = & -  \frac{1}{3}\, ({\bf \nabla \cdot u})\: p\, - b_\ell(p) 
- D_{p }\,\frac{\partial \ln f_j}{\partial p}. \label{pdot:eq}
\end{eqnarray}
In writing Eq.~(\ref{dcei:eq})-(\ref{fpe:eq}) we have emphasized that
both advection and diffusion terms along the momentum coordinate can
be cast in conservative form, in analogy to the corresponding terms in
configuration space.  This allows us to adopt a Godunov-like scheme
for the numerical integration of those terms.  However, we place them
on the right hand side of the Eq.~(\ref{dcei:eq}) because they will
effectively be treated as source terms of $n_{p_j}$.  Finally, on the
right hand side of Eq.~(\ref{dcei:eq}), we have added a source term,
$J_j$, which represents the rate of production of CR particles due to
the diffusive shock acceleration mechanism.  We do this only {\it pro
  forma}: because we always use Glimm's method to advance shocks, the
Riemann solver effectively subsumes the role of the term $J_j$. In
other words, $J_j$ will not be treated as an explicit source term but
as an implicit part of the shock solution computed through the Riemann
solver.  Thus, it will be sufficient for our purposes to only
formulate a prescription for the postshock values of the set
$\{n_{p_j}\}$, which we do in Section~\ref{crmrs:se}.
\subsection{Fluxes in Momentum Space}
Time integration of Eq.~(\ref{dcei:eq}) due to the fluxes in momentum
space is done following the method proposed in~\cite{min01}.  We can
also retain the diffusive term although here we limit the discussion to
the case of a small diffusion coefficient, $D_p$, which can be treated
explicitly, i.e.
\begin{equation} 
\tau_{D_p} \equiv \frac{\Delta p^2}{D_p} \gg \Delta t .
\end{equation}
Here $\tau_{D_p}$ is the characteristic diffusion time in momentum
space, $\Delta p=p_{j+\frac{1}{2}} - p_{j-\frac{1}{2}}$, the momentum
bin size, and $\Delta t$ the time step.  Then, the terms on the
right-hand-side of Eq.~(\ref{dcei:eq}) produce a change in $n_{p_j}$
such that
\begin{equation} \label{dnpadv:eq}
n^{t+\Delta t}_{p_j} - n^t_{p_j} = - \frac{\Delta t}{\Delta p}
 \, \left( F_{p_{j+\frac{1}{2}}}^{n+\frac{1}{2}} - F_{p_{j-\frac{1}{2}}}^{n+\frac{1}{2}} \right) -
\Delta t \, [({\bf \nabla }_x \cdot F_x) +  J_j] ,
\end{equation}
where the superscript $n+\frac{1}{2}$ indicates time centering and the
last term on the right hand side will be specified in Sections
\ref{glimm:se} and \ref{ggm:se}.  We can start estimating the
time-averaged flux in momentum space at time $t=n\Delta t$
as~\cite{min01}
\begin{equation}  \label{fluxp:eq}
F_{p_{j+\frac{1}{2}}}^n  = 
\frac{\Delta p}{\Delta t} \int_{p_{j+\frac{1}{2}}}^{p_u} 4\pi \;p^2\; f^n_{j_u}(p)\; dp , 
\end{equation}
which is obtained by time integrating Eq.~(\ref{fpe:eq}) and by
changing integration variable from time to momentum~\cite{min01}.  In
Eq.~(\ref{fluxp:eq}) $f^n_{j_u}$ is the upstream distribution function
at time $t$ defined at the grid point in momentum space
\begin{equation} \label{jdef:eq}
j_u = 
   \left\{ \begin{array}{lll}  
   j+1 & \mbox{if} & \dot{p}_{j+\frac{1}{2}} \le 0 ,\\ 
   j   & \mbox{if} & \dot{p}_{j+\frac{1}{2}}  >  0 ,
   \end{array} \right.
\end{equation}
and $p_u$ is the upstream momentum, solution of the integral equation
\begin{equation}   \label{pu:eq}
\Delta t = \int^{p_u}_{p_{j+\frac{1}{2}}} 
\frac{dp }{\dot{p}}.
\end{equation}
Note that the denominator of the above integrand function has
typically a polynomial form so that the integral can be computed in
closed form~\cite{min01}.  Finally, a time centered estimate of the
term, $\tilde\nabla F^{n+\frac{1}{2}}_p$, is obtained by taking the
average of the fluxes, $F_p$, computed at $t$ and $t+\Delta t$, as
usually done for nonstiff source terms.  Time centering is relevant
because the function in Eq.~(\ref{pu:eq}) in general depends on the
local properties of the fluid, such as density and temperature, which
change with time.  Note that there are two ways to compute the
divergence of the velocity in the $\dot p$ term: with a cell centered
scheme $(\nabla u)_i = (u_{i+1}-u_{i-1})/2\Delta x$ or with a
staggered scheme, $(\nabla u)_i =
(u_{i+\frac{1}{2}}-u_{i-\frac{1}{2}})/\Delta x$ in which case $u_{i\pm
  \frac{1}{2}}$ is computed as part of the solution to the Riemann
problem that one has to solve in order to advance the fluid equations
and estimate the spatial terms appearing in Eq.~(\ref{dnpadv:eq}).
This is described in the next section.  In the following we always use
the staggered scheme except in the Godunov's predictor step.

\section{Riemann Problem for Cosmic-ray Hydrodynamics} 
\label{rpcrh:se}

In this section we describe the modifications to the Riemann problem
in the presence of CRs. Without loss of generality we restrict to the
one-dimensional case. In addition, for the sake of clarity, in the
following discussion we shall neglect the spatial diffusion term,
except for the fact that it is implicitly assumed to be at work at
shocks, where it causes CR particles to be accelerated.
In smooth flows this term is assumed to be slow and is
taken into account with an explicit conservative formulation.

We begin with rewriting our system of equations in conservative form:
\begin{equation} \label{husys:eq}
\frac{\partial U}{\partial t} + \nabla_x \cdot F_x  = S(U),
\end{equation}
where
\begin{equation} \label{usys:eq}
U \equiv
   \left( \begin{array}{l} \rho \\ \rho u\\ \rho e  \\ n_{p_j} 
    \end{array} \right),
 F_{x} \equiv
   \left( \begin{array}{c} \rho u \\
   \rho u^2+P \\
   (\rho e+P) u \\ 
    n_{p_j}u
    \end{array} 
    \right),
 S \equiv
   \left( \begin{array}{c} 0 \\ 0 \\  \dot{E}_{loss} \\ 
    \tilde\nabla_p F_{p_j} + J_{p_j} 
      \end{array}
      \right), 
\end{equation}
the entry $n_{p_j}$ is repeated for all momentum bins, e.g. for $j=0,N_p-1$,
and we have introduced the total pressure and specific energy as
\begin{eqnarray}
P & = & P_g+P_c , \\
e & = & \frac{1}{2} u^2 + \frac{P_g}{\rho (\gamma_g-1)} 
+ \frac{P_c}{\rho (\gamma_c-1)} .
\end{eqnarray}

Losses in the total energy of the system are due to escape of 
energetic particles and, in principle, radiative losses. Thus we write
\begin{equation} \label{eloss:eq}
\dot{E}_{loss} = - 4\pi m_c c^2 \left\{\left.
p^2 f \dot{p} \left[(p^2+1)^{\frac{1}{2}}-1\right] \right|_{p=p_{max}} \!\!
+ \int_{p_{min}}^{p_{max}} b_{r} \frac{fp^3}{(p^2 + 1)^{\frac{1}{2}}} \, dp
\right\},
\end{equation}
with $b_{r}$ referring to radiative losses only. Note that the first
term is different from zero only if $\dot p>0$ because we assume
$f(p)=0$ for $p>p_{max}$.  In addition, just like the term $J_j$
discussed in the previous section, the source term $\dot{E}_{loss}$ 
will not be treated explicitly but will be part of the shock 
solution computed through the Riemann solver.

The conservative character of the Eq.~(\ref{husys:eq}) 
for the system in Eq.~(\ref{usys:eq}) suggests that the jump conditions
across a shock wave can be written in the usual way, provided that the
total, thermal plus cosmic-ray, energy and pressure are used.
However, there is an additional complication related to the way in
which the total pressure and energy is partitioned between CRs and
thermal components. This is further addressed in
Section~\ref{crmns:se}.

For the sake of clarity in the ensuing discussion, we now outline the
Riemann problem.  Suppose that at $t=0$ the gas is described by
\begin{equation} \label{riemann:eq}
U = 
\left\{ \begin{array}{lll}
    U^r & \mbox{if} &  x \ge 0, \\ 
    U^l & \mbox{if} &  x < 0.
         \end{array} \right.
\end{equation}
The solution at $t>0$ is in general characterized by two waves: a
backward moving wave separating the states $(U^l,U^{*l})$ and a
forward moving wave separating the states $(U^r,U^{*r})$. Each wave
will be either a shock wave or a rarefaction wave. The central states
$U^{*l},U^{*r}$ are separated by a slip line across which the velocity
and total pressure $(u^*,P^*)$ remain constant, but the density and
individual pressure components,
$(\rho^{*\oo},P^{*\oo}_g,P^{*\oo}_c),~\oo=l,r$, in general will
change.  The value of these three quantities in each intermediate
state, $U^{*\oo}$, will be reconstructed from $U^{\oo}$ and the type
of wave connecting the two states. In the following two subsections we
describe the structure of such waves.

\subsection{Characteristic Analysis} \label{charan:se}

Transforming into primitive variable space we obtain
\begin{equation}
  U \equiv
  \left( \begin{array}{l} \rho \\ \rho u \\ \rho e \\ n_{p_j} \end{array} \right) \;   \longrightarrow  V \equiv
  \left( \begin{array}{l} \rho \\ u \\ P_g \\ P_c \\ y_{p_j} \end{array} \right),
\end{equation}
where we have introduced the CR concentrations, $y_{p_j} \equiv
n_{p_j}m_p/\rho $.  These quantities are followed as passive scalars
and, for simplicity, will be omitted in the following analysis.  Note,
however, that the source term ensures that there is consistency
between their evolution and that of $P_c$, to which they are related
through Eq.~(\ref{ppl:eq}), (\ref{kave:eq}) and (\ref{pc:eq}).

The system of equations for the primitive variables, $V$, is obtained
with the following transformation
\begin{eqnarray} \label{hwsys:eq}
&& \frac{\partial V}{\partial t} - A(V) \frac{\partial V}{\partial x}  = S_V, \\
A(V)  &  = &  \nabla _U V \cdot \nabla _U F \cdot \nabla _V U,~~~S_V  = \nabla _U V \cdot S,
\end{eqnarray}
where
\begin{eqnarray} \label{aarray:eq}
A & = &
\left (
\begin{array}{ccccccc}
 u &  \rho       &  0      & 0        \\
 0 &  u          &  1/\rho & 1/\rho  \\
 0 & \rho c_g^2  &  u      & 0     \\
 0 & \rho c_c^2  &  0      & u     \\
\end{array}
\right), \\ \label{sspeeds:eq}
c_g \equiv \left(\frac{\gamma_g P_g}{\rho}\right)^{\frac{1}{2}},\!\!\!\!\!
&& c_c\equiv \left(\frac{\gamma_c P_c}{\rho} \right)^{\frac{1}{2}},~~
c_s\equiv  \left(\frac{\gamma_g P_g+\gamma_c P_c}{\rho}\right)^{\frac{1}{2}}.
\end{eqnarray}
Here $c_g$ and $c_c$ correspond to the speed of sound associated with
the gas and CR pressure respectively.  (In principle, a coefficient
$(\partial \ln P_c/\partial\ln \rho)_s$ should be used instead of
$\gamma_c$ in the definition of $c_g$, but the difference is
negligible for the purpose of this paper.)  Solving for ${\rm
  Det}(A-\lambda {\rm I})=0$, the eigenvalues of $A$ are found to be
$\lambda_0=u-c_s,~\lambda_{1}=u,~ \lambda_{2}=u,~\lambda_{3}=u+c_s$.
The associated left and right eigenvalues are, respectively,
\begin{eqnarray}
L =
\left (
\begin{array}{cccccc}
0 & -\rho/2c_s &  1/2c_s^2   &  1/2c_s^2 \\
1 & 0          & -1/ c_s^2   & -1/ c_s^2  \\
0 & 0          & c_c^2/c_s^2 & -c_g^2/c_s^2  \\
0 & \rho/2c_s  &  1/ 2c_s^2  &  1/2c_s^2 \\
\end{array}
\right) 
\end{eqnarray}
and
\begin{eqnarray}
R =
\left (
\begin{array}{cccccc}
1         & 1 & 0 & 1 \\
-c_s/\rho & 0 & 0 & c_s/\rho \\
c_g^2     & 0 & 1 &  c_g^2 \\
c_c^2     & 0 & -1 & c_c^2 \\
\end{array}
\right).
\end{eqnarray}

According to this analysis `simple waves', including rarefaction waves,
are described by the following equations
\begin{eqnarray}
\frac{d\rho}{dP} & = & \frac{1}{C_s^{2}}, \label{drdp:eq} \\
\frac{du}{dP} & = & \pm \frac{1}{C_s},  \label{dudp:eq} \\
\frac{dP_c}{dP_g} & = & \frac{C_c^2}{C_g^2}, \label{dpcdpg:eq} 
\end{eqnarray}
where $C_\# =\rho c_\#,\#=g,c,s$, is the Lagrangian sound speed or
mass flux across the wave. The first two
equations~(\ref{drdp:eq})-(\ref{dudp:eq}) are the usual relation for
hydrodynamics with the sound speed modified to account for the CR
pressure. The last equation describes the change of CR pressure as a
function of the gas pressure during an adiabatic process.

\subsection{Cosmic-ray Mediated Numerical Shocks} \label{crmns:se}

In the following we consider the structure of a shock modified by the
presence of accelerated CR particles. As in the case of hydrodynamics,
we assume that steady state conditions have been reached.  When CR
acceleration operates this takes of order the timescale to accelerate
thermal particles to relativistic energies. While substantially longer
than for a pure hydrodynamic case, in our astrophysical limit this is
typically still much shorter than the size of a time step (cf.~\cite{kajo07}
for further details on this).

If we label quantities far upstream and far downstream of the shock
with, $-$ and $+$, respectively, the Rankine-Hugoniot conditions read
\begin{eqnarray}
u^+ - u^-  & =  &\pm \frac{P^+ - P^-}{W},  \label{ujmp:eq} \\
\frac{P^+ - P^-}{W^2}  & = & - \left(\frac{1}{\rho^+} - \frac{1}{\rho^-} \right), \label{rjmp:eq}  \\
\rho^+ e^+ - \rho^- e^-  & = & - \frac{1}{2}\,\frac{P^+ + P^-}{1/\rho^+ - 1/\rho^-}, 
\label{ejmp:eq}
\end{eqnarray}
where $W$ is the Lagrangian speed of propagation of nonlinear waves.
These do not yet specify the amount of CR energy and pressure
generated by the dissipation mechanism. In fact, $P_c$ and $\gamma_c$
are not and cannot be specified by the above equations alone.  To
do that we use the following facts without proof (but
see~\cite{drvo81,axlemk82,acblpe84}).  In the presence of CRs the
shock discontinuity is replaced by a precursor, where the gas is
compressed adiabatically by the upstreaming CRs and $P_g\propto
\rho^\gamma \propto u^{-\gamma}$; the precursor is immediately
followed by a viscous subshock where entropy is generated but both the
CR pressure and CR energy flux remain continuous.  This means that the
structure of the subshock is purely hydrodynamical.
If we use the label, $0$, to indicate the quantities just prior to the
subshock, the compression at the precursor is given by
\begin{equation} \label{precursor:eq}
r_p = \frac{u^-}{u^0}.
\end{equation}
Define the shock Mach number as
\begin{equation}
\mach \equiv \frac{u^-}{c^-_g}.
\end{equation}
The gas pressure jump across the total shock transition, being the
result of the precursor adiabatic compression and subshock
compression\footnote{A term describing non adiabatic heating in the
  precursor due, in particular, to Alfv\'en wave dissipation, can in
  principle also be included in Eq.~(\ref{pg+:eq}). It is neglected here,
  however, for simplicity.}, reads~\cite{acblpe84}
\begin{equation} \label{pg+:eq}
\frac{P^+_g}{P^-_g} = \frac{2\gamma_g}{\gamma_g+1} \,\frac{\mach ^2}{r_p}
 -  \frac{\gamma_g-1}{\gamma_g+1} \, r_p^{\gamma_g}.
\end{equation}
An analogous relation can be obtained for the conservation of mass equation,
namely
\begin{equation}
\frac{u^-}{u^+} = 
r_p \frac{\gamma_g+1} {\gamma_g-1+2 r_p^{\gamma_g+1}\mach^{-2}}.
\end{equation}
Using these results with Euler equation gives~\cite{acblpe84}
\begin{equation} \label{pc+:eq}
\frac{P^+_c}{P^-_g}  =
\frac{P^-_c}{P^-_g} + 1 - r_p^{\gamma_g} + \gamma_g(1-r_p^{-1}) \mach^2.
\end{equation}
Recalling that, $\mach = W/C$, Eq.~(\ref{pg+:eq}) and~(\ref{pc+:eq})
then lead to the following modified definition of the Lagrangian speed
of nonlinear waves
\begin{equation} \label{lspeed:eq}
W = C_g^- \, \left[ \frac{2 r_p^{\gamma_g}}{\gamma_g +1 - r_p^{-1} (\gamma_g -1)} \,
\left( 1 + \frac{\gamma_g+1}{2\gamma_g\,r_p^{\gamma_g} }\frac {P^+ - P^-}{P_g^-} \right) \right]^{\frac{1}{2}} ,
\end{equation}
where $C_g^-=(\gamma_g \rho^- P_g^- )^{\frac{1}{2}},~ P=P_g+P_c$.  Finally,
the value of the CR adiabatic index downstream of the shock,
$\gamma_c^+$, is determined by the energy equation through the
following relations:
\begin{eqnarray}
\label{gc+:eq}
\gamma_c^+ & = & \frac{\gamma_g g^+}{\gamma_g g^+ - \gamma_g +1} , \\
 g^+ & = & \frac{g^- \frac{P^-_c}{P^-_g} + 1-r_p^{\gamma_g-1} + 
 \frac{\gamma_g -1}{2}(1-r_p^{-2}) \mach^2+ 
 \frac{\gamma_g-1}{\gamma_g P_g^-u^-}Q_{loss}}{\left( \frac{\gamma_g-1}{\gamma_g+1} r_p^{-1} + \frac{2}{\gamma_g+1} \frac{r_p^{\gamma_g}}{\mach^2} \right)
\, \left[\frac{P^-_c}{P^-_g} + 1-r_p^{\gamma_g} + \gamma_g (1-r_p^{-1}) \mach^2 \right] }, \\
g^- & = & \frac{\gamma_c^- (\gamma_g-1) }{ \gamma_g(\gamma_c^--1)}, \\ \label{qloss:eq}
Q_{loss} &=& \int_{x^-}^{x^+} \dot{E}_{loss} dx,
\end{eqnarray}
where $\dot E_{loss}$ is given in Eq.~(\ref{eloss:eq}). The term
$Q_{loss}\le0$ describes the energy losses occurring at the precursor
and shock front.  It becomes important when the CR acceleration
efficiency is very high.  In this case, this term must also be
specified consistently with the acceleration efficiency by the kinetic
solution.
\subsection{Riemann Solver Procedure} \label{crmrs:se}
Having specified the form of the rarefaction and compression waves
modified by the CRs, we can now define the procedure for solving the
Riemann problem.  First note that, provided the shock acceleration
efficiency, $\eta(U^-,\mach)$, as a function of the upstream
conditions ($U^-$) and the shock Mach number ($\mach$), we solve
Eq.~(\ref{pc+:eq}) to derive a similar function for the compression at
the precursor,
\begin{equation}
r_p = r_p(U^-,\mach).
\label{rp:eq}
\end{equation}

Given the left and right states in Eq.~(\ref{riemann:eq}), we then
want to compute the intersection point, $(u^*,P^*)$, of the two wave
curves passing through $U^{l},U^{r}$ in the $P\!-\!u$ plane. For this
we use the iterative technique proposed in~\cite{vanleer79}, with the
two shock approximations~\cite{colella82} and additional modifications
which we describe next.

In the absence of shocks we change the Lagrangian speed of nonlinear waves
as follows in order to account for the CR pressure
\begin{equation}
W = C^{l,r}_s 
\left( 1+ \frac{\gamma_g+1}{2\gamma_g}\frac{P^*-P^{l,r}}{P^{l,r}}\right)^{\frac{1}{2}},
\label{csnd:eq}
\end{equation}
with $C^{l,r}_s = \rho^{l,r} c_s^{l,r}$.  If a shock is present we
instead use $W$ given in Eq.~(\ref{lspeed:eq}) with $P^+$ replaced by
$P^*$.  However, unlike the pure hydrodynamic case, $W$ now also
depends on $r_p(\mach)$. Thus, using $\mach=W/C_g$, we have the
implicit equation
\begin{equation}
W = W\left[P^*,r_p\left(\frac{W}{C_g}\right)\right],
\label{w:eq}
\end{equation}
which also needs to be solved iteratively.  In addition the tangent
slopes to the wave curves in the $P\!-\!u$ plane, which are used in
the iterative procedure to find $P^*$, are modified according to
\begin{eqnarray}
Z \equiv \left| \frac{dP^*}{du^*} \right| & = & 
\frac{2\,W^3}{W^2+ C^2} \longrightarrow \frac{2\,W^3}{W^2+ Y C^2}, \\
Y & = &  \frac{2\,r_p^{\gamma_g}}{\gamma_g+1-(\gamma_g-1) r_p^{-1}}.
\end{eqnarray}

To summarize, the iteration procedure is now given by :
\begin{algorithmic}
\STATE
\STATE
{$\nu=0,\quad P^{*}_0 = 
[C_s^lP^r + C_s^rP^l - C_s^lC_s^r (u^r-u^l)]/(C_s^l + C_s^r)$}
\WHILE {not converged} 
{
\STATE{$\nu$++}
\IF {$P^*_{\nu-1} > P^{l,r}$}\STATE
  {$W^{l,r}_\nu = W^{l,r}_\nu(P^*_{\nu-1},W^{l,r}_\nu/C_g^{l,r})$}\STATE
  {$\mach^{l,r} = W^{l,r}_\nu/C_g^{l,r}$}\STATE
  {$r_p^{l,r} = r_p(U^{l,r},\mach^{l,r}),\quad Y^{l,r} = Y(r_p^{l,r}),\quad C^{l,r} = C^{l,r}_g $}
\ELSE\STATE
  { $W^{l,r}_\nu = W^{l,r}_\nu(P^*_{\nu-1})$} \STATE
  {$r^{l,r}_p = 1,\quad Y^{l,r} = 1,\quad C^{l,r} = C^{l,r}_s $}
\ENDIF\STATE
{$Z^{l,r} = Z^{l,r}(W^{l,r}_\nu,C^{l,r},Y^{l,r}) $}\STATE
{$u^{*,l} = u^l - (P^*_{\nu-1}-P^l)/W^l,\quad u^{*,r} = u^r - (P^*_{\nu-1}-P^r)/W^r$} \STATE
{$P^{*}_{\nu} = P^*_{\nu-1} - [Z^lZ^r/(Z^l+Z^r)] \, (u^{*,r}-u^{*,l}) $}
}
\ENDWHILE
\STATE
{$u^{*} = (W_{\nu}^lu^l + W_{\nu}^ru^r + p^l-p^r)/(W_{\nu}^l + W_{\nu}^r)$}
\STATE
\end{algorithmic}
The criterion for convergence can be set to be, $|P^*_\nu -
P^*_{\nu-1}|/P^*_\nu < \epsilon$, where $\epsilon$ is a parameter
that sets the error tolerance.  Note that at the end of the above
procedure, in addition to $P^*$ and $u^*$, we have also solved for
$W^{l,r}$ and, therefore, $\mach^{l,r}$ and $r_p^{l,r}$.

Once the left and right moving waves have been determined, we proceed
as follows~\cite{colella82}.  In searching for the solution at a given
point, $\xi\equiv x/t$, we set $\sigma= {\rm sign} (\xi -u^*)$ and
define
\begin{eqnarray}
& & (U^\oo,W^\oo,c^\oo,\mach^\oo,r_p^\oo,\gamma_c^\oo) = 
\left\{ \begin{array}{lll}
    (U^l,W^l,c^l,\mach^l,r_p^l,\gamma_c^l) & \mbox{if} &  \sigma > 0,  \\
    (U^r,W^r,c^r,\mach^r,r_p^r,\gamma_c^r) & \mbox{if} &  \sigma < 0, 
         \end{array} \right. \\
& &          \hat u^\oo = \sigma u^\oo, \quad  \hat \xi = \sigma \xi, \quad   \hat u^* = \sigma u^* .
\end{eqnarray}
We then complete the definition of the intermediate state $U^{*\oo}$.
If the latter is separated from $U^\oo$ through a rarefaction wave,
knowing $P^*,P_g^\oo$ and $P_c^\oo$ we can use Eq.~(\ref{dpcdpg:eq})
to estimate $P_g^{*\oo},~P_c^{*\oo}$.  This amounts to solving for
$P_g^{*\oo}$ the nonlinear equation
\begin{equation}\label{pgrar:eq}
P^*  =  P_g^{*\oo} + P_c^\oo \,
  \left( \frac{P_g^{*\oo}}{P_g^\oo} \right)^{ \gamma_c^\oo/\gamma_g } , 
\end{equation}
and then setting
\begin{equation}
P_c^{*\oo}  =  P^* - P_g^{*\oo}. \label{pcstar:eq}
\end{equation}
In Eq.~(\ref{pgrar:eq}), $\gamma_c^\oo$ is the CR adiabatic index of
the $U^\oo$ state which remains unchanged during an adiabatic process.
The density is then estimated through the polytropic law
\begin{equation}\label{rhorar:eq}
\rho^{*\oo}=\rho^\oo \,\left(\frac{P_g^{*\oo}}{P_g^\oo}\right)^{1/\gamma_g}
= \rho^\oo \,\left(\frac{P_c^{*\oo}}{P_c^\oo}\right)^{1/\gamma_c^o},
\end{equation}
Finally, $n^{*\oo}_{p_i}= n^{\oo}_{p_i} (\rho^{*\oo}/\rho^\oo)$.
(Note that, in Eq.~(\ref{pgrar:eq}) and~(\ref{rhorar:eq}), as in the
definition of the sound speeds, Eq.~(\ref{sspeeds:eq}), the quantity
$(\partial \ln P_c/\partial \ln\rho)_s$ should be used as adiabatic
index.)

In the case of a shock wave, on the other hand, knowing both $r_p^\oo$
and $\mach^\oo$, we estimate $P_g^{*\oo}$ with Eq.~(\ref{pg+:eq}),
$P_c^{*\oo}$ with Eq.~(\ref{pcstar:eq}) and $\rho^{*\oo}$ with
Eq.~(\ref{rjmp:eq}).  In addition, $\gamma_c^{*\oo}$ is defined by
Eq.~(\ref{gc+:eq}) and, with $f^+(p;U^\oo,\mach^\oo)$ specified by the
input kinetic-model, the downstream number density of CRs in each bin
is given by
\begin{equation}
\label{npc+:eq}
n^{*\oo}_{p_j}=\int_{p_{j-\frac{1}{2}}}^{p_{j+\frac{1}{2}}} 4\pi\,p^2 \;f^+(p;U^\oo,\mach^\oo)\; dp.
\end{equation}
Note that the consistency between $\eta(U^\oo,\mach^\oo)$,
$f^+(p;U^\oo,\mach^\oo)$ and $Q_{loss}$
ensures consistency between the values of
$P_C^{*\oo},~\gamma_c^{*\oo}$ and $n^{*\oo}_{p_j}$ also.

We then evaluate $c_s^*$ through Eq.~(\ref{sspeeds:eq}) and define the
wave speeds
\begin{equation}
\hat\lambda^\oo,~\hat\lambda^* =
\left\{ \begin{array}{lll}
    \hat u^\oo + c^\oo,~\hat u^*+c^*  & \mbox{if} & P^*<P^\oo, \\ 
     \hat u^\oo +\frac{W^\oo}{\rho^\oo} & \mbox{if} &  P^* \ge P^\oo. \\ 
         \end{array} \right. 
\end{equation}
If $\xi$ lies ahead or behind the o-wave we can set the solution to
\begin{equation}
\rho, u, P_g, P_c, n_{p_i} = 
\left\{ \begin{array}{lll}
    \rho^{*\oo}, u^{*\oo}, P_g^{*\oo}, P_c^{*\oo},n_{p_i}^{*\oo}  & \mbox{if} & \hat \xi \le \hat \lambda ^*, \\ 
    \rho^\oo, u^\oo, P_g^\oo, P_c^\oo,n_{p_i}^\oo    & \mbox{if} &  \hat \xi \ge \hat \lambda ^\oo.  \\ 
         \end{array} \right. 
\end{equation}
However, if $\hat \lambda ^* < \xi < \hat \lambda ^\oo$, we have to
evaluate the solution inside a rarefaction wave.  This requires
integration of the system~(\ref{drdp:eq})-(\ref{dpcdpg:eq}), which
cannot be done in closed form and can be expensive. An alternative
method is to linearly interpolate between the states $U^\oo$ and
$U^{*\oo}$ as
\begin{eqnarray} \label{rwfint:eq}
U & = & \zeta \, U^{*\oo} + (1-\zeta) \, U^\oo, \\
\zeta & = & \frac{\hat\lambda^\oo - \xi}{\hat\lambda^\oo-\hat\lambda^*}.
\end{eqnarray}
This works just fine for Godunov's method. However, we find that for
Glimm's method, when strong rarefactions are involved, it is important
to use the exact approach in order to avoid spurious effects.

Before concluding the description of the Riemann solver we point out
the presence of a slight inconsistency in the formulation. In fact
$\lim_{r_p\rightarrow 1} W \neq C_s$, i.e. the speed of weakly
nonlinear waves does not tend to the speed of sound waves. This is a
consequence of the large gap in physical scales between the sound
waves driven by the total CR and thermal pressure and the CR mediated
shock waves. The former are in fact long wavelength perturbations on
which scale the diffusion is slow and unimportant. Such scales are
much larger than those characterizing the structure of a shock.  Thus
the conflict in trying to reconcile the two solutions is due to the
impossibility of following the intermediate scales.  Since the
nonlinear effects due to the process of CR shock acceleration are only
important for strong shocks, a natural way of solving the above
conflict is to assume that a shock solution is adopted only if the
shock propagation speed exceeds that of the sound speed given in
Eq.~(\ref{csnd:eq}). This leads to the following condition
\begin{equation} \label{minpcjmp:eq}
\Delta P > \frac{\gamma_g+1}{2\gamma_g}\frac{\gamma_c}{\gamma_g} P_c.
\end{equation}

\section{Implementations}

The method described in the previous section for the solution of the
Riemann problem can be employed to construct time dependent solutions
to the system of equations (\ref{rhoe:eq})-(\ref{dce:eq}) on a grid.

\subsection{Glimm's Method} \label{glimm:se}

The first implementation we describe is based on Glimm's method.
Following~\cite{colella82}, here we simply outline the main procedure
that we use and refer the reader to the original
references~\cite{glimm65,chorin76,chorin77,colella82} for a detailed
description.

Consider a piecewise constant approximate solution at time $t^n=n\Delta t$
\begin{equation}
U(x,t^n) = U^n_i , \quad
\left(i-\frac{1}{2}\right)\Delta x \le x < \left(i+\frac{1}{2}\right)\Delta x,
\quad i\in {\mathcal D},
\end{equation}
where $\Delta x$ is the mesh size, $\Delta t$ the timestep and
${\mathcal D}$ the computational domain.  We seek to advance the
solution by one timestep to $t=(n+1)\Delta t$. To do that we solve the
Riemann problem at each cell interface, $i-\frac{1}{2}$, with left and right
states given by $U^n_{i-1}$ and $U^n_{i}$.  Denote the solution with
\begin{equation} \label{ivp:eq}
{\mathcal R}_{i-\frac{1}{2},n} \left[\frac{x-(i-\frac{1}{2})\Delta x}{t-n\Delta t} \right],
\end{equation}
where we have made explicit use of its property of self-similarity.
If the choice of the timestep is sufficiently small, say
\begin{eqnarray}
\Delta t & < & \frac{1}{2} \,\frac{\Delta x}{\lambda_{\max}},\\
\lambda_{\max} & = & \max (|u^n_i| + c^n_{si}), \forall i \in {\mathcal D},
\end{eqnarray}
the wave solutions of the Riemann problems at each cell interface will
not interact with each other. Then the set of Riemann solutions,
$\{{\mathcal R}_{i-\frac{1}{2},n}, i\in {\mathcal D}\}$, each covering a
region, $(i-1)\Delta x \le x < i\Delta x$, defines an exact solution,
$U^{e,n}(x,t)$, to the initial value problem in~(\ref{ivp:eq}) for the
time interval, $t^n<t\le t^{n+1}$.  The solution at each grid point,
$i$, and time, $t^{n+1}$, is obtained by random sampling $U^{e,n}$ as
follows: evaluate the solution at the point, $x=(i-1+a^{n+1})\Delta
x$, within the region covered by ${\mathcal R}_{i-\frac{1}{2},n}$, where,
$a^{n+1}$, is a randomly chosen number, $a^{n+1}\in [0,1)$.  Note that
$x\in i$ if $a^{n+1}\ge \frac{1}{2}$ and $x\in (i-1)$ if $a^{n+1} < \frac{1}{2}$.  We
can then define
\begin{equation} \label{glimmsol:eq}
(U^{n+1}_i)_{\rm Glimm}= 
\left\{ \begin{array}{lll}
    {\mathcal R}_{i-\frac{1}{2},n} 
    \left[\left(a^{n+1}-\frac{1}{2}\right) \frac{\Delta x}{\Delta t} \right] & 
    \mbox{if} &  a^{n+1} > \frac{1}{2}, \\ 
    {\mathcal R}_{i+\frac{1}{2},n}  
    \left[\left(a^{n+1}-\frac{1}{2}\right) \frac{\Delta x}{\Delta t} \right] &
    \mbox{if} &  a^{n+1} \le \frac{1}{2}.
         \end{array} \right.
\end{equation}
Following~\cite{colella82}, we use a sampling procedure that is based
on van der Corput's pseudo-random sequence, so that $a^n$ is the
$n_{th}$ element of that sequence.

\subsection{Hybrid Glimm-Godunov's Method} \label{ggm:se}

Shock waves are the only features that need to be propagated without
numerical smearing. Therefore, we have also implemented a hybrid
scheme which uses Glimm's method to advance shock fronts and Godunov's
method for smooth parts of the flow.  Having described Glimm's method
in the previous section, here we briefly outline a scheme based on the
higher order Godunov's method.

In Godunov's method the solution is updated with a conservative scheme
\begin{equation}\label{godup:eq}
(U^{n+1}_i)_{\rm Godunov} = U^n_i -
\frac{\Delta t}{\Delta x} 
\left(F^{n+\frac{1}{2}}_{i+\frac{1}{2}}-F^{n+\frac{1}{2}}_{i-\frac{1}{2}} \right) + 
\Delta t\, S(U^{n+\frac{1}{2}}),
\end{equation}
where the source term has been described in Section~\ref{dms:se}.

The fluxes at the cell faces are given by
\begin{equation}
F^{n+\frac{1}{2}}_{i+\frac{1}{2}} = F\left(V^{n+\frac{1}{2}}_{i+\frac{1}{2}}\right),
\end{equation}
where $V^{n+\frac{1}{2}}_{i+\frac{1}{2}}$ is obtained by solving the
Riemann problem (discussed extensively in Section~\ref{rpcrh:se}) with
left and right states $\left(V_{i,+} , V_{i+1,-}\right)$. These states
correspond to up-wind time averages, which allow to achieve second
order accuracy.  They are reconstructed from the cell center, taking
into account the effects of spatial gradients and the source term, as
follows.  At each grid point, $i$, we compute centered and one-sided
slopes and use van Leer's limiter to make the final choice about the
local slope, $\Delta V_i$.  Then, the up-wind, time averaged left
($-$) and right ($+$) states at cell faces are
\begin{eqnarray}
V_{i, \pm}  & =  & V_i ^n+ \frac{1}{2} 
\left(\pm I - \frac{ \Delta t}{\Delta x} \, A \right) \, 
{\mathcal P}_\pm (\Delta V_i) , \\
\label{finalpred:eq}
V_{i, \pm}  & =  & 
V_{ i, \pm}  + \frac{\Delta t}{2} \, S_{Vi}^n .
\end{eqnarray}
Here $A$ is given in Eq.~(\ref{aarray:eq}), $I$ is the identity
operator and $n$ indicates the time-step corresponding to time $t$. In
addition
\begin{equation}
{\mathcal P}_\pm (V) = \sum _{\pm \lambda_j >0} (l_j \cdot V) \cdot r_j
\end{equation}
projects out from the state $V$ the components carried by
characteristics that propagate away from the cell interface ($l_j,
r_j$ are the left and right eigenstates respectively and $\lambda_j$
is the corresponding eigenvalue, described in
Section~\ref{charan:se}).

In conclusion, our hybrid scheme can be summarized as
\begin{equation} \label{hybridsol:eq}
U^{n+1}_i = 
\left\{ \begin{array}{ll}
     (U^{n+1}_i)_{\rm Glimm} & \mbox{if $i$ is shocked}, \\ 
    (U^{n+1}_i)_{\rm Godunov} &  \mbox{otherwise.}   
         \end{array} \right.
\end{equation}
We say that the cell, $i$, is {\it shocked} if a shock is going to
cross it during the next timestep. In order for this to happen at
least a shock moving with speed $u_s$ with respect to the grid must be
present at the interface, $i-{\rm sign}(u_s) \frac{1}{2}$. The criterion for
deciding whether or not a wave across an interface qualifies as a
shock will be based on the strength of the pressure jump across it,
$|P_{i+1}-P_i|/\min(P_{i+1},P_i)$, and shall take the
condition~(\ref{minpcjmp:eq}) into account.

\section{Tests} \label{tests:se}

We now present a few tests illustrating the performance of the methods
described in the previous sections.  The tests consist of a set of
Riemann problems with initial conditions
\begin{equation} \label{ic:eq}
(\rho, u,P_g,P_c,\gamma_c, f(p))  [x,t=0] = 
   \left\{ \begin{array}{lll}  
   (\rho^l, u^l, P_g^l, P_c^l, \gamma_c^l, f^l(p)) & \mbox{if} &  x\le 0.5 ,\\ 
   (\rho^r, u^r, P_g^r, P_c^r, \gamma_c^r, f^r(p)) &  \mbox{if} &  x>0.5,
    \end{array} \right. 
\end{equation}
for which we compare the numerical and the `exact' solutions. The
`exact' solution is obtained by solving the Riemann problem as
outlined in Section~\ref{rpcrh:se}, numerically but without any of the
approximations involved in the Riemann solvers for the numerical
methods.  In particular, no two shock approximation is made, and the
exact expression for the speed of rarefaction waves is used.

In order to allow for an easier comparison with the `exact' solution
we only retain the adiabatic terms in the diffusion-convection
equation, i.e.  we neglect the terms $D_p$ in Eq.(\ref{fpe:eq}) and
$b_\ell(p)$ in Eq.(\ref{pdot:eq}).  Similarly we let momentum space
range over 15 orders of magnitude with $p_{min}=10^{-5}$ and
$p_{max}=10^{10}$. This choice, while unrealistic, is made in order to
minimize energy losses due to fluxes across boundaries in momentum
space when studying rarefaction waves.

For simplicity the initial conditions for the CR distribution function
are specified as an unbroken power-law,
\begin{equation} \label{pl:eq}
f^{l,r}(p)=f^{l,r}_0(p/p_0)^{-q^{l,r}},
\end{equation}
with $p_{min}\le p\le p_{max}$.
CR particles return to the thermal pool for $p \le p_{min}$
and escape the system for $p\ge p_{max}$.  Due to the lack of energy
losses the evolution of $f(p)$, followed with
the scheme presented in Section~\ref{dms:se}, becomes trivial.
However, the accuracy of that scheme
has already been extensively tested in~\cite{min01}, so that
here we focus solely on the quality of the hydrodynamic solutions.

In solving the Riemann test problems with the numerical methods
presented in this paper we always employ a grid of 128 mesh points on
a domain of size unity (so that $\Delta x = 1/128$). We use
$\gamma_g=5/3$.  In addition we use $N_p=16$ momentum bins.
Throughout the Section time is expressed in adimensional code units.

As already mentioned, when evaluating the solution inside a
rarefaction wave with the Riemann solver, we have a choice of either
integrating directly the Eq.~(\ref{drdp:eq})-(\ref{dpcdpg:eq}) or use
the approximate Eq.~(\ref{rwfint:eq}). When using Glimm's method we
test both approaches and compare the results, whereas when using
Godunov's method we only employ the approximate approach which turns
out sufficiently accurate.

For shock waves, we adopt the following simple prescription defining
the shock acceleration efficiency
\begin{eqnarray}
\eta(U^-,\mach)=
\left\{ \begin{array}{ll}
{\mathcal A} \, 
\left[1-\exp\left(-\frac{\mach-\mach_{min}}{\mach_s} \right) \right], 
    & \mbox{if}~ \mach>\mach_{min}, \\ 
    0 & \mbox{otherwise} ,
       \end{array} \right.
\end{eqnarray}
in which the fraction of total momentum impinging on the shock and
dissipated into CR pressure depends solely on the shock Mach number.
In the above expression, $\mach_{min}$ and $\mach_{s}$ are a threshold
and scale parameter, respectively.  While clearly a simplification,
the functional form of $\eta$, with a sharp rise for
$\mach>\mach_{min}$, followed by a flattening for $\mach>\mach_{s}$,
is partially inspired by thermal leakage models and the numerical
results described in~\cite{kajo05}.  We take ${\mathcal A} = 0.8$,
$\mach_s = 5.77$ and $\mach_{min}=1.5$.  In this simplified model we
use values of $Q_{loss}\le0$ that, based on the prescribed
acceleration efficiency and the energy equation, allow for $\gamma_c
\ge 4/3$. With the above choices, the resulting shock solutions always
admit a subshock, i.e. completely smooth shock transitions do not
appear.

The accelerated CR distribution function is assumed to be an unbroken
power-law as in Eq.~(\ref{pl:eq}).  Using
Eq.~(\ref{pc:eq})-(\ref{gc:eq}), we can thus write the following
relation between the slope and the CR adiabatic index
\begin{eqnarray} \label{plslope:eq}
q & = & 3\,(1+\alpha[q] [\gamma_c-1]), \\
\alpha(q)  & = & 1 - \frac{4\pi m_cc^2}{3P_c(q)}
\left\{p^3\, f(p)\,\left[(p^2+1)^{\frac{1}{2}}-1\right]\right\}_{p_{min}}^{p_{max}},
\end{eqnarray}
in which $\alpha(q)\ge0$ is a function of $q$ through $P_c(q)$ and $f(p)$. 
The above relations imply that $q$ takes a value in the
interval $(3,\infty)$ as $\gamma_c$ ranges between $4/3$ and $5/3$.

It should be pointed out that in a realistic shock the slope of the
distribution function, $q$, in general depends on $p$ and is
determined self-consistently with the velocity profile in the
precursor and subshock. However, the simplification made here
about the distribution function, as well as other naive assumptions
made earlier in this section, are solely for the sake
of simplicity or easy comparison with exact solutions. 
Nothing prevents the use of more sophisticated and
realistic kinetic models for  $\eta(\mach),\,p_{min},\,p_{max},\,q(p)$
with the numerical method presented in this paper.

\subsection{Shocks}  \label{sam.se}
\begin{figure} 
\begin{center}
\includegraphics[height=0.315\textheight, scale=1.0]{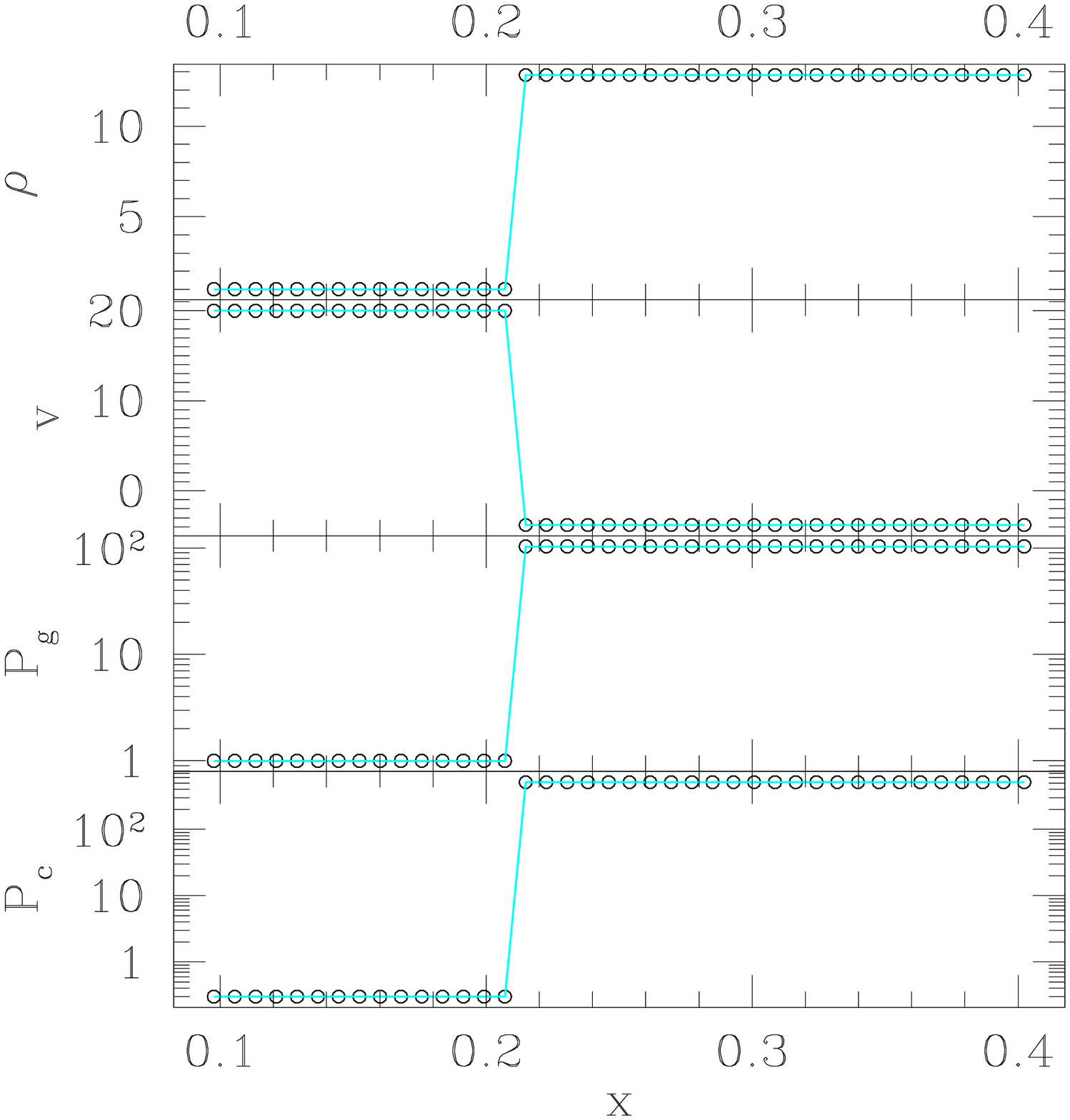}\includegraphics[height=0.315\textheight, scale=1.0]{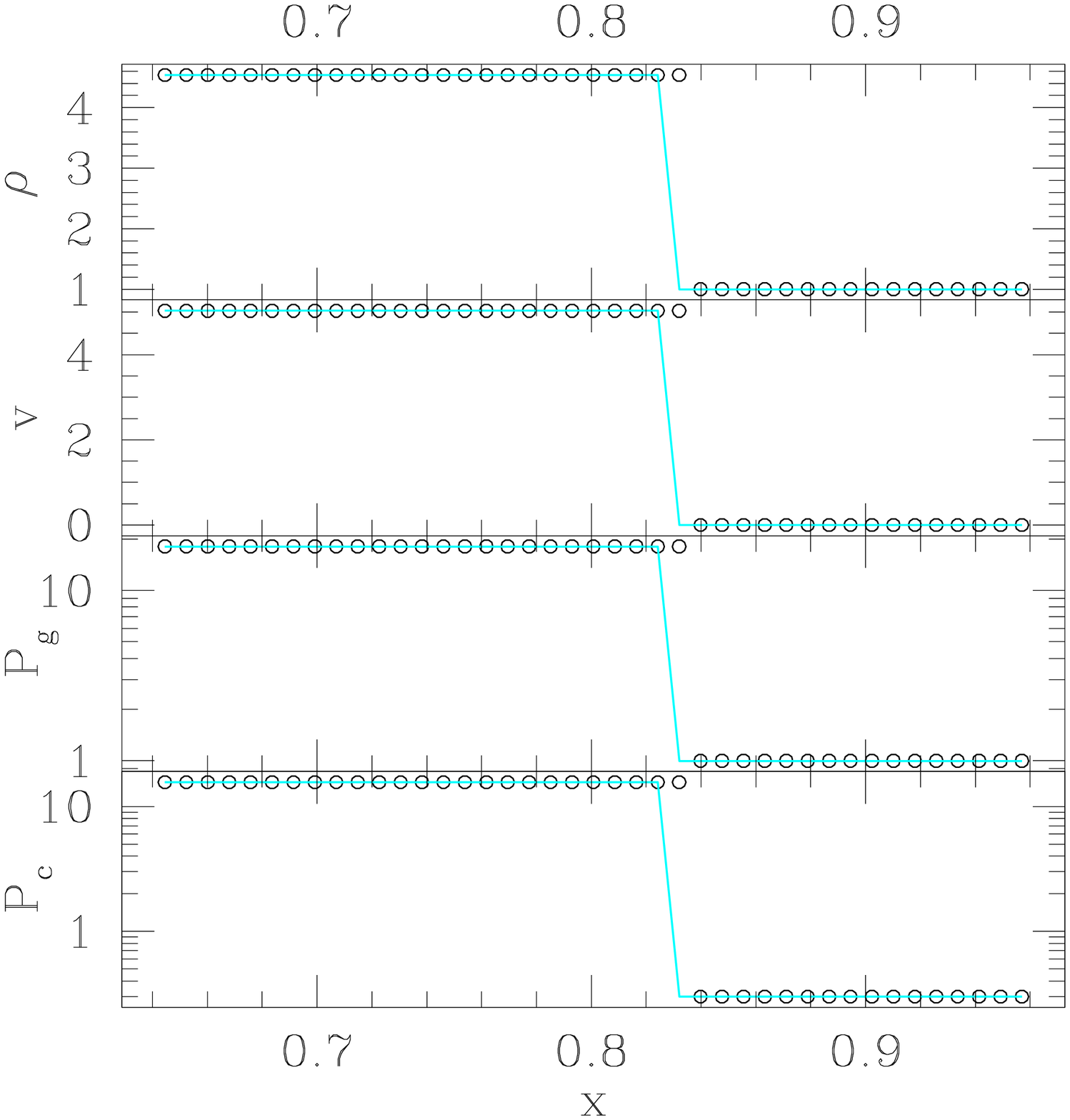}
\caption{Open symbols: numerical
  solutions obtained with Glimm's method for two shock
  problems.  Solid line: `exact' solution. 
  Left: solution at $t=0.05$ for a left moving shock with 
  $\mach=20$ (initial conditions in Eq.~[\ref{sl:eq}]).
  Right: solution at $t=0.051$ for a left moving shock with $\mach=5$ 
  (initial conditions in Eq.~[\ref{sr:eq}]).
\label{shocks:fig}}
\end{center}
\end{figure}
We begin with three shock problems. Two shocks with mild upstream 
ratio of CR to thermal pressure, one moving leftward with
$\mach = 20$ and the other moving rightward with
$\mach = 5$. And a shock with 
upstream ratio of CR to thermal pressure equal to
one, and moving leftward with $\mach = 10$.
The problems are specified by the following parameters:
\begin{equation} \label{sl:eq}
\begin{array}{lll}  
Q_{loss} = -155.3796, & Q_{loss}/F^-(\rho e) = 0.461045724 &  \\
\rho^l = 1.0, & u^l=20, & P_g^l  = 1.0, \\
P_c^l  = 0.3, & \gamma_c^l = 1.34, &   \\
\rho^r = 12.8305315, & u^r=-3.80751023  , & P_g^r = 103.280692, \\
P_c^r = 512.726578, & \gamma_c^r =1.33433,  & 
\end{array} 
\end{equation}
for the first problem, 
\begin{equation} \label{sr:eq}
\begin{array}{lll}  
Q_{loss} = 0.0, & Q_{loss}/F^-(\rho e) = 0.0 &  \\
\rho^l = 4.53983646, & u^l=5.03312097, & P_g^l  = 18.1559788, \\
P_c^l  = 15.6326773, & \gamma_c^l = 1.34218, &   \\
\rho^r = 1.0, & u^r=0.0  , & P_g^r = 1.0, \\
P_c^r = 0.3, & \gamma_c^r =1.34.  & 
\end{array} 
\end{equation}
for the second problem, and 
\begin{equation} \label{shp-:eq}
\begin{array}{lll}
Q_{loss} = -19.93444, & Q_{loss}/F^-(\rho e) = 0.222258137 &  \\
\rho^l = 1.0, & u^l=10, & P_g^l  = 1.0, \\
P_c^l  = 1.0, & \gamma_c^l = 1.35, &   \\
\rho^r = 7.64274954, & u^r=-1.22076909, & P_g^r = 42.8536158, \\
P_c^r = 104.00589, & \gamma_c^r =1.33433,  &   
\end{array} 
\end{equation}
\begin{figure} 
\begin{center}
\includegraphics[height=0.4\textheight, scale=1.0]{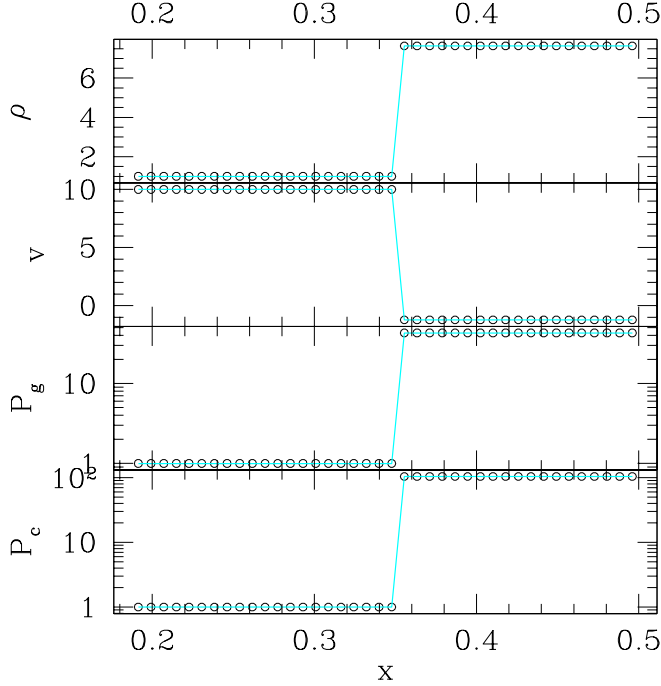}
\caption{Numerical
  solutions obtained with Glimm's method (open symbols) and
 `exact' solution (solid line) at $t=0.05$ for for a left moving shock
with $\mach=10$ and upstream CR to thermal pressure ratio equal to one
(initial conditions in Eq.~[\ref{shp-:eq}]).
\label{highpc-:fig}}
\end{center}
\end{figure}
for the third problem. The initial CR distribution functions are
specified by Eq.~(\ref{pl:eq}), with slope $q^{l,r}$ determined
respectively by $\gamma_c^{l,r}$ through Eq.~(\ref{plslope:eq}).  In
addition to the initial left/right states of the Riemann problems we
have also specified the parameter $Q_{loss}$ defined in
Eq.~(\ref{qloss:eq}) and its ratio to the upstream energy flux,
$F^-(\rho e)$. So, in the above three examples about 46\%, 0\% and
22\%, respectively, of the energy flux through the shock front is
carried away by escaping CR particles with $p>p_{max}$.

In these tests the role of Godunov's method in the hybrid formulation
is trivial. Therefore we only show the results obtained with Glimm's
method.  The left and right moving shocks described by
Eq.~(\ref{sl:eq})-(\ref{sr:eq}) are presented in the left and right
panels of Fig.~\ref{shocks:fig}, respectively, while Fig.
\ref{highpc-:fig} refers to the case described by Eq.~(\ref{shp-:eq}).
All plots correspond to a solution time, $t=0.05$. For each plot the four
panels show, from top to bottom, gas density, velocity, gas pressure
and CR pressure.  The numerical solution reproduces the `exact'
solution very well, without oscillations or artifacts, despite the
fact that the CR pressure is comparable or significantly higher than
the thermal pressure.  Note that the front of the right moving shock
is displaced with respect to the `exact' solution by one cell.  This
is a characteristic of Glimm's method: as it advances the shock front
in discrete steps of size $\Delta x$, it may inevitably place the
shock on a grid position that is offset with respect to the `true'
position.  By using van der Corput sequence, however, the offset is at
most one zone and it eventually becomes negligible when compared to
the distance traveled by the shock~\cite{colella82}.

\subsection{Rarefactions}
\begin{figure} 
\begin{center}
\includegraphics[height=0.315\textheight, scale=1.0]{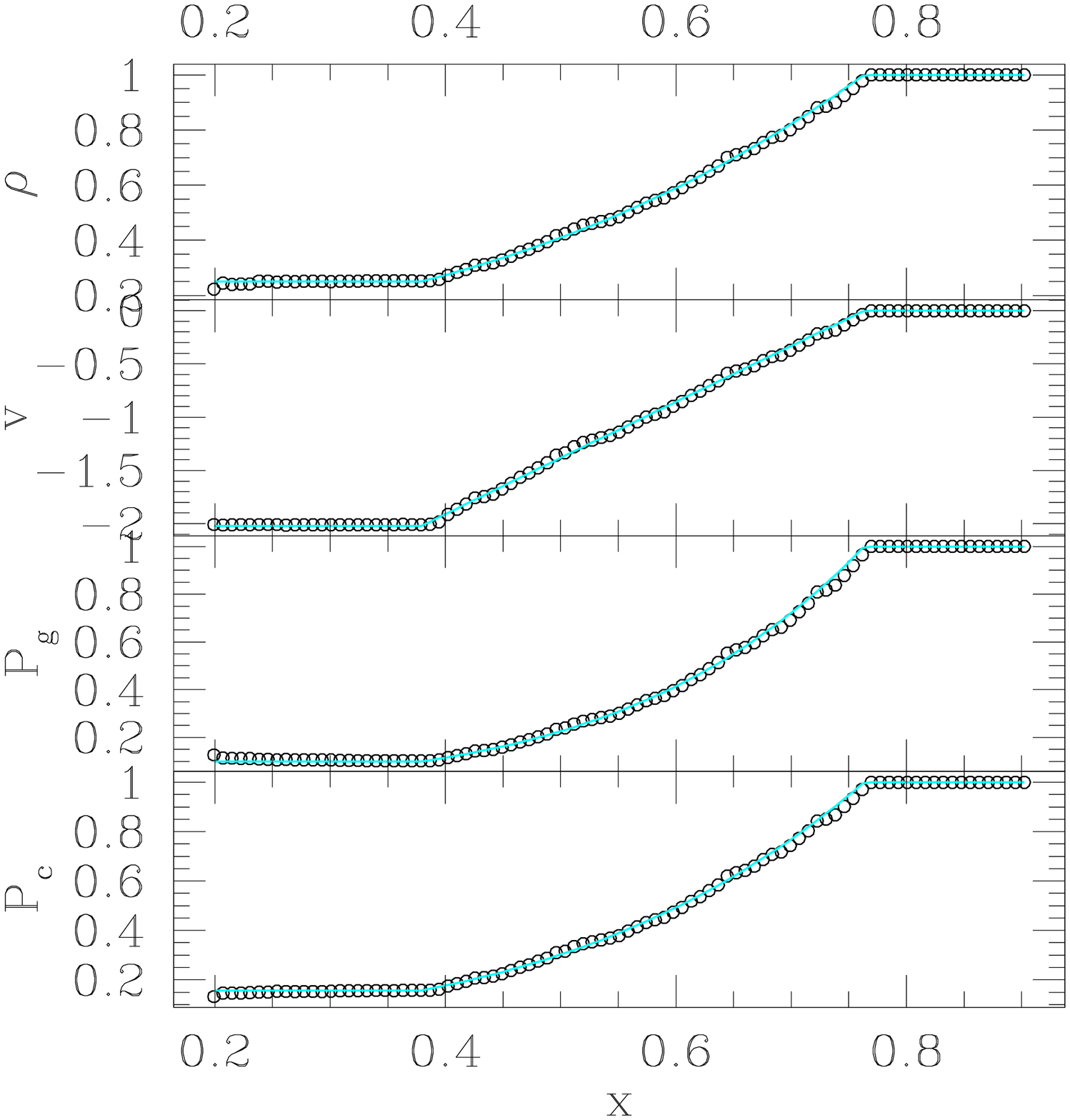}\includegraphics[height=0.315\textheight, scale=1.0]{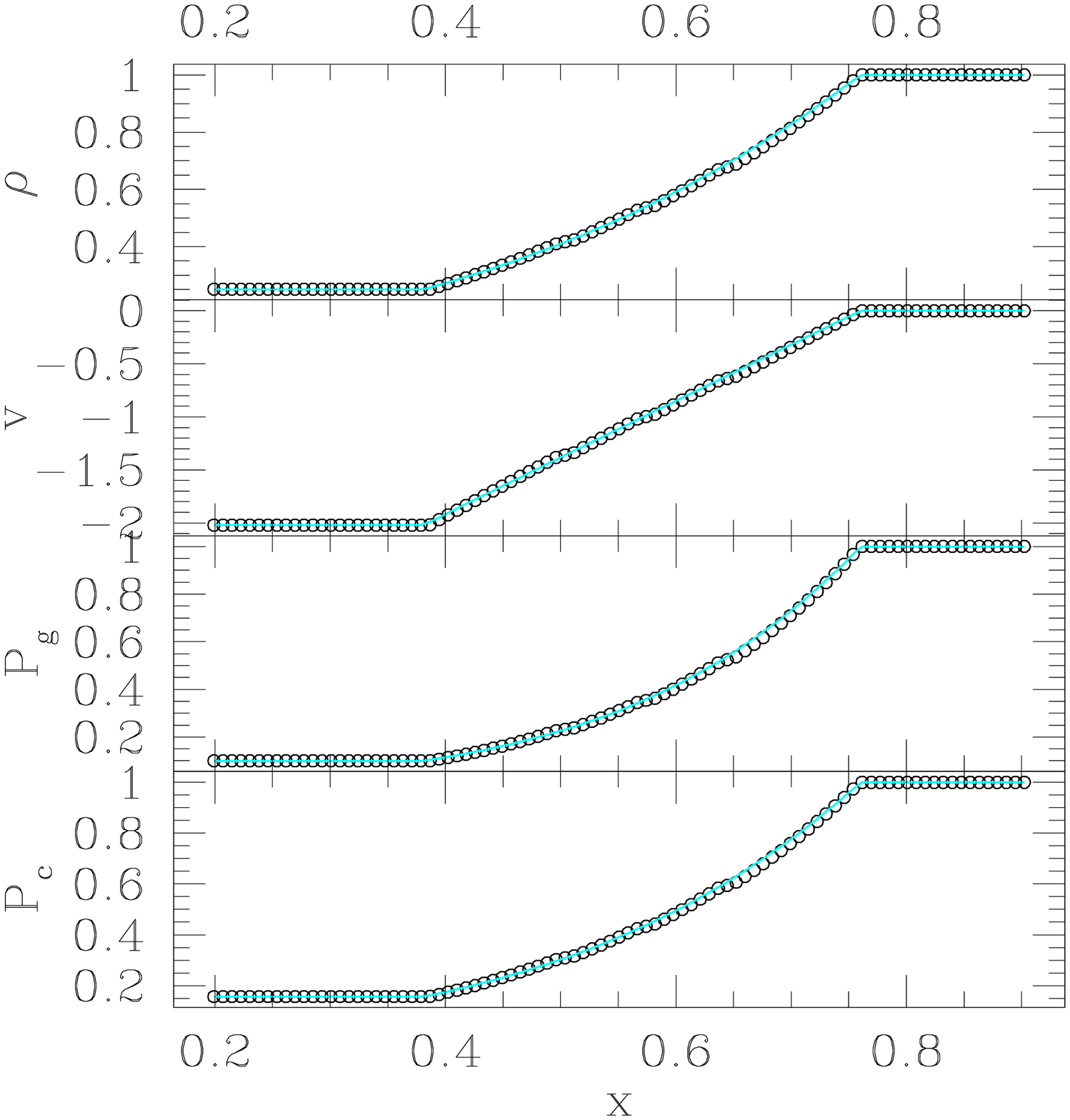}
\caption{Open symbols: rarefaction wave solutions at, $t=0.1504683$, obtained with Glimm's 
method using the approximate (left) and exact (right) 
rarefaction solution in the Riemann solver.  Solid line: 
The solid lines indicates 
the exact solutions. The initial conditions are described in 
Eq.~(\ref{rwic:eq})
and correspond to a rarefaction wave in the $\lambda_+$ characteristic family.
\label{raref:fig}}
\end{center}
\end{figure}
We now turn to the following initial value problem:
\begin{equation} \label{rwic:eq}
\begin{array}{lll} 
\rho^l = 0.251188643, & u^l=-2.01959396, & P_g^l  = 0.1, \\
P_c^l  = 0.15703628, & \gamma_c^l = 1.34, &   \\
\rho^r = 1.0, & u^r=0.0, & P_g^r = 1.0, \\
P_c^r = 1.0, & \gamma_c^r =1.34,  & 
\end{array} 
\end{equation}
representing a rarefaction wave in the $\lambda_+$ characteristic
family. The initial CR distribution functions 
are power-laws, Eq.~(\ref{pl:eq}), with the same slope $q$ determined
by $\gamma_c^l=\gamma_c^r$ through Eq.~(\ref{plslope:eq}). 
Note that in this case $Q_{loss}=0$. 
Fig.~\ref{raref:fig} shows the solution at time $t=0.15$,
obtained with Glimm's method using the approximate (left) and
exact (right) rarefaction solution in the Riemann solver,
respectively.  Again, from top to bottom, each panel shows gas
density, velocity, gas pressure and CR pressure.

Glimm's solutions appear slightly ragged.  The raggedness in the right
panel is purely due to the sampling procedure and does not correspond
to an oscillatory behavior.  In particular, despite their appearance,
the points on the rarefaction curve remain connected through the correct
wave solution and the ragged character may or may not be there
depending on the specific sample that is being drawn. We show an
example of this in the next test, where Glimm's solution of a shock
tube problem is characterized by a smooth rarefaction wave.

On the other hand, the solution on the left panel shows additional
irregularity which is due to the approximations in Riemann solver.  In
particular the approximations involved in evaluating the solution
inside a rarefaction wave add spurious structure to the Riemann
solution which is picked up by Glimm's method. This is undesirable
because some of the spurious features may be amplified and even become
unstable.

\begin{figure} 
\begin{center}
\includegraphics[height=0.315\textheight, scale=1.0]{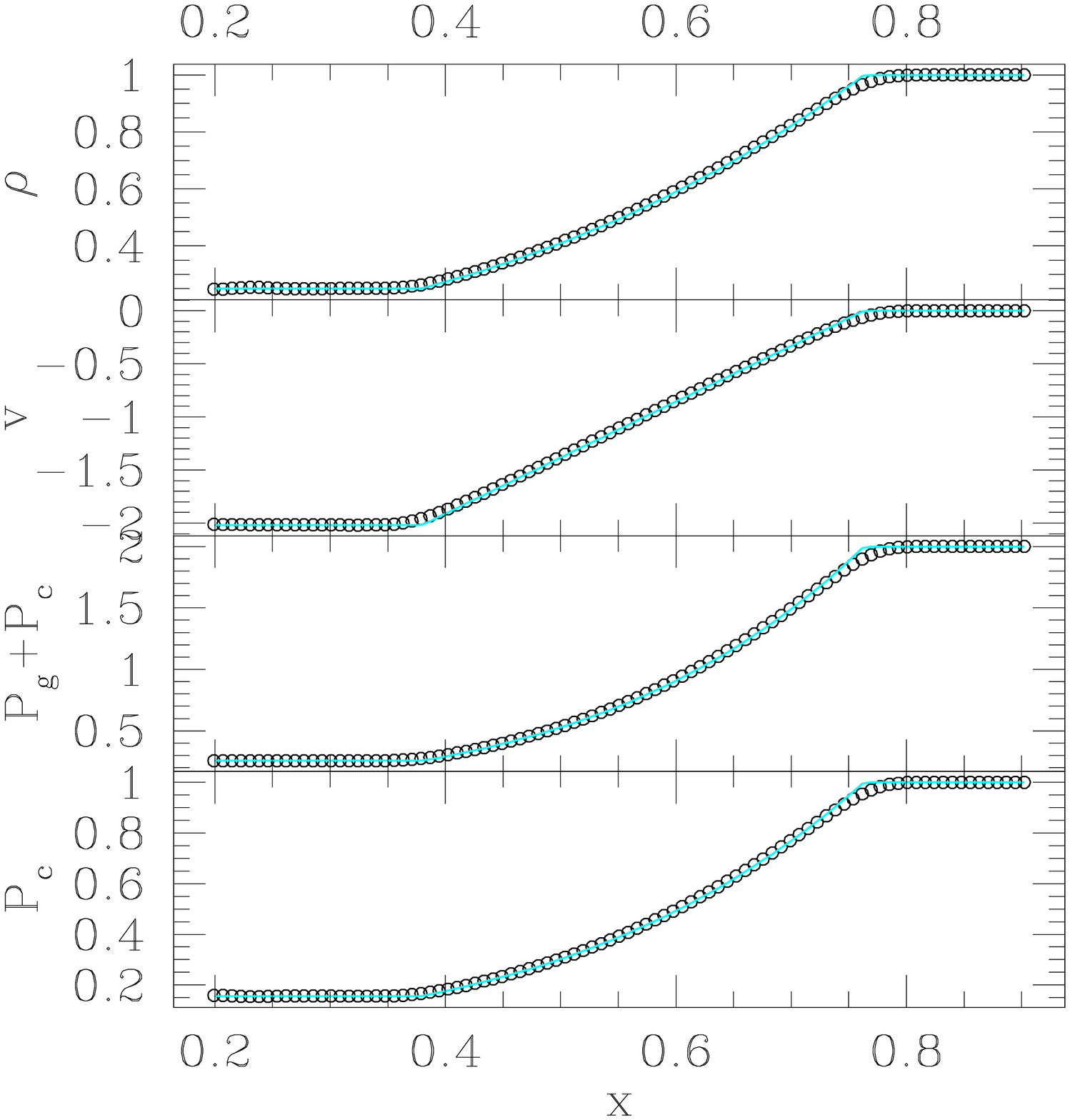}\includegraphics[height=0.315\textheight, scale=1.0]{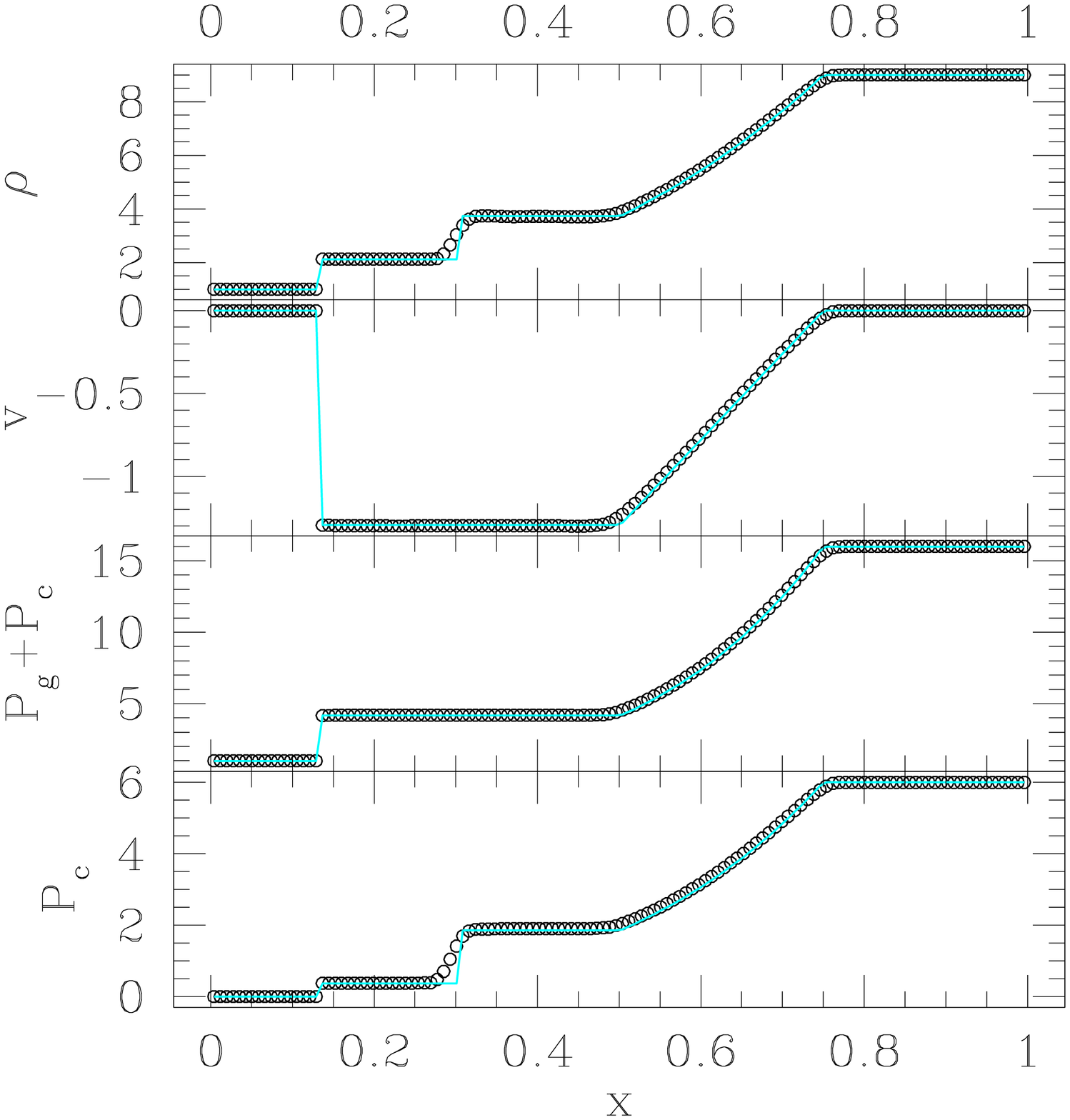}
\caption{Open symbols: numerical solutions obtained with the hybrid Glimm-Godunov's method.
Solid line: exact solution.
Left panel: rarefaction wave at, $t=0.1504683$. The initial conditions
are described in Eq.~(\ref{rwic:eq}, cf. Fig.~\ref{raref:fig}).
Right panel: shock tube problem at, $t=0.150794$.
The initial conditions are described in Eq.~(\ref{st:eq}).
\label{god:fig}}
\end{center}
\end{figure}

Finally, the left panel of Fig.~\ref{god:fig} shows the solution
obtained with Godunov's method for the same rarefaction wave problem.
Overall the numerical solution reproduces accurately the evolution of
the rarefaction wave. Compared with Glimm's method, the solution is
now smooth across the wave, although it appears slightly less sharp at
the head of the wave.
\subsection{Shock Tubes}
We conclude the set of tests with a shock tube problem with
the following initial conditions 
\begin{equation} \label{st:eq}
\begin{array}{lllll} 
\rho^l = 1.0, & u^l=0.0, & P_g^l = 1.0, & P_c^l = 0.0, & (\gamma_c^l),   \\
\rho^r = 9.0, & u^r=0.0, & P_g^r = 10, & P_c^r = 6.0, & \gamma_c^r =1.33433.
\end{array} 
\end{equation}
The initial CR distribution function for the right state
is a power-law, Eq.~(\ref{pl:eq}), with slope $q^{r}$ determined
by $\gamma_c^{r}$ through Eq.~(\ref{plslope:eq}).
Note that given the null value of the CR pressure on the left state,
we need not specify $\gamma_c^l$ nor $f^l(p)$.  In the shock tube problem in
general a shock, a rarefaction and a contact discontinuity develop.
In this case the shock is weak and $Q_{loss}=0$.

\begin{figure} 
\begin{center}
\includegraphics[height=0.315\textheight, scale=1.0]{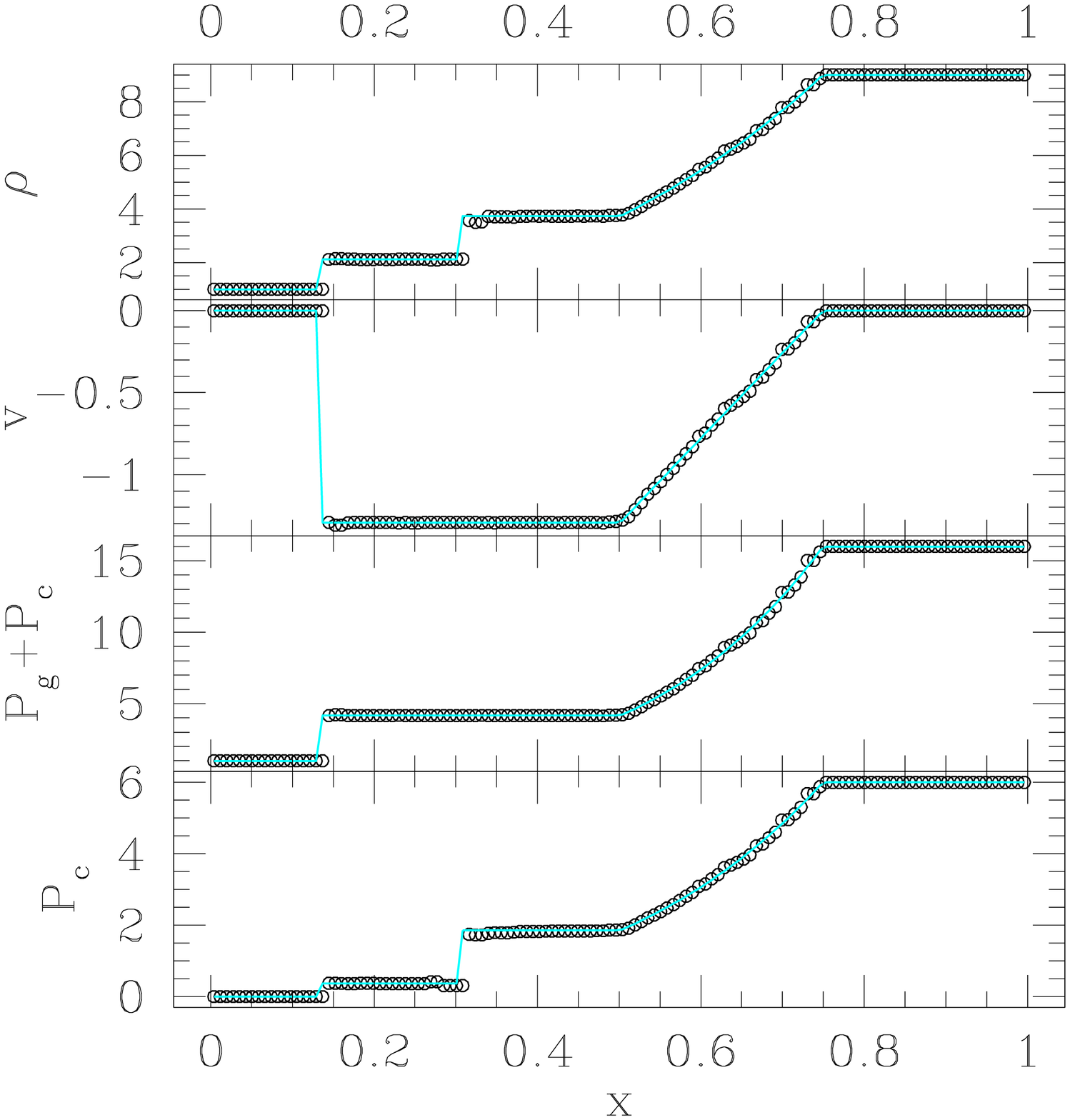}\includegraphics[height=0.315\textheight, scale=1.0]{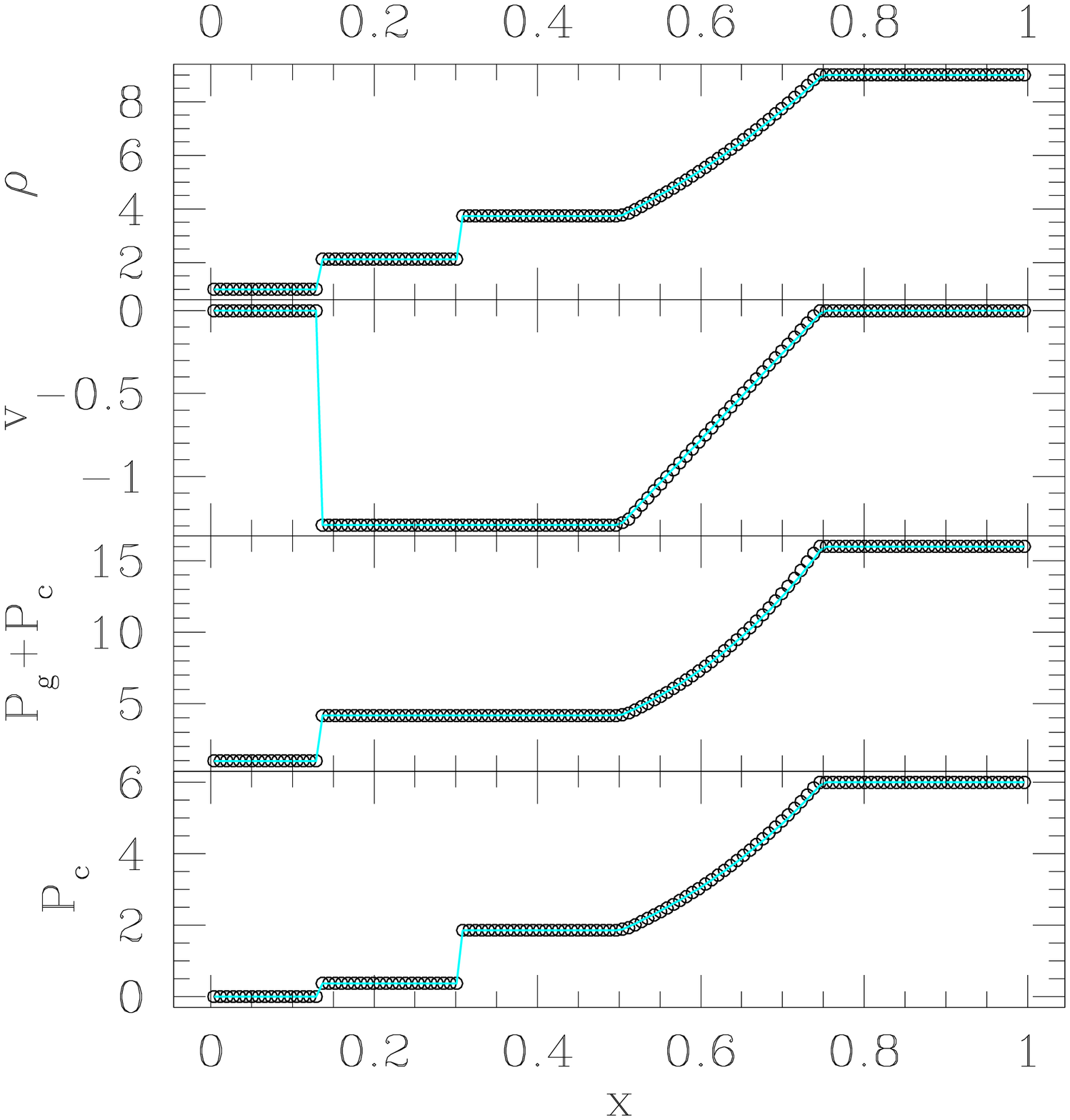}
\caption{  Opens symbols:
shock tube problem solution at, $t=0.150794$, 
obtained with Glimm's method, 
using the approximate (left) and
exact (right) rarefaction solution in the Riemann solver.
Solid lines: `exact' solution.
The initial conditions are described in Eq.~(\ref{st:eq}, 
cf. Fig.~\ref{god:fig}). 
\label{st_glimm:fig}}
\end{center}
\end{figure}

The numerical results at $t=0.15$ are shown in the right panel of
Fig.~\ref{god:fig} for the hybrid Glimm-Godunov's method and in
Fig.~\ref{st_glimm:fig} for Glimm's method. As in the previous test,
the latter figure shows both the results obtained by employing an
approximate (left) and exact (right) rarefaction solution in the
Riemann solver. Note that from top to bottom, each panel now shows gas
density, velocity, total (gas+CR) pressure and CR pressure,
respectively.

The left panel of Fig.~\ref{st_glimm:fig} shows additional evidence
that when using the approximate rarefaction solution with Glimm's
method, spurious structure appears in the numerical solution.  This
affects not only the rarefaction wave but also the intermediate state.
When the exact rarefaction solution is employed however, Glimm's
method produces a highly accurate result. Unlike the earlier test, the
rarefaction curve is now smooth, confirming how the raggedness in the
previous test was due to the sampling procedure. Notice how Glimm's
method preserves the sharpness of both the shock and contact
discontinuity.  As already pointed out, however, their position may be
displaced with respect to the exact solution by at most one grid cell.

The right panel of Fig.~\ref{god:fig} shows that the numerical
solution produced by the hybrid method is also highly accurate.  As a
sanity check we notice that in switching between Glimm and Godunov's
formulations no spurious effect is introduced.  In addition, the shock
position is the same as in Glimm's method solution.  The rarefaction
part of the solution is well captured, although now
the foot of the rarefaction is not as sharp as in Glimm's
method. Finally, the contact discontinuity spreads over a few cells,
which is characteristic of Godunov's method.  Note that the contact
discontinuity appears both in the density and in the pressure
components, $P_g$ and $P_c$. This, however, does not affect the total
pressure $P_g+P_c$, which remains perfectly constant in the
intermediate state between the rarefaction and the shock, guaranteeing
a correct velocity profile as well.

\section{Conclusions} \label{concl:se}

In this paper we have developed a method to include the dynamical
effects of CR pressure on a hydrodynamical system.

In smooth flows this is achieved by modification of the characteristic
equations which define the waves that carry information in the
CR-hydro fluid. The exchange of energy between the two CR particles
and the fluid component as a result of diffusive processes both in
configuration and momentum space, is modeled with a flux conserving
method.

Regarding the solution at shock waves, we have shown that once the
acceleration efficiency has been specified as a function of the
upstream conditions and shock Mach number, the shock CR mediation and
the induced substructure can be correctly taken into account in the
fluid solution by modifying the procedure for the Riemann solver.

We have implemented two numerical schemes for obtaining time dependent
solutions on a computational grid. One based on Glimm's method and
another based on a hybrid Glimm-Godunov method.  In both approaches we
exploit the ability of Glimm's method to preserve the discontinuous
character of shocks.  This is useful because when combined with the
aforementioned modified Riemann solver it provides a natural scheme
for advancing the shock solution at the correct speed, meaning that
the CR dynamical effects are taken into account without resolving the
shock substructure.

In smooth flows Glimm's method is not the only possible choice and
Godunov's method can also be employed.  Our tests show that Glimm's
method is rather sensitive to the approximations assumed in the
Riemann solver procedure.  In particular, when evaluating the solution
inside a rarefaction wave it is important to solve the exact
equations~(\ref{drdp:eq})-(\ref{dpcdpg:eq}) or else spurious features
will appear. This is not the case with Godunov's method. Compared to
the version of Glimm's method with the exact rarefaction solution in
the Riemann solver, Godunov's produces smoother solutions and is only
slightly less sharp in capturing the head of rarefaction waves. So the
hybrid method is also a viable option.

The proposed method can be readily employed for the study of
one-dimensional models of astrophysical systems, such as simplified
radially symmetric descriptions of, e.g., Supernova Remnants or Galaxy
Clusters. The potential benefit of this scheme, however, lies in the
possibility of coupling it with three-dimensional shock tracking
algorithms or extending Glimm's method to three dimensions.  In this
case the dynamical role of CRs in astrophysical systems can be studied
without geometrical restrictions.  Since numerical dissipation is
necessary for stable hydrodynamic shocks in three
dimensions~\cite{colella82}, but spoils the correct CR-hydrodynamic
shock solution, shock tracking algorithms may offer the only way to
self-consistently study CR-hydrodynamics in multi-dimensions.

\vskip 1truecm 
\leftline{\bf Acknowledgment} 

I am grateful to Phil Colella for valuable discussion and to Tom Jones
for useful comments.  I acknowledge support by the Swiss Institute of
Technology through a Zwicky Prize Fellowship.

\bibliographystyle{plain}
\bibliography{papers,books,proceed,codes}

\end{document}

%% file: macrosAPS.tex
\def\lsim{\raise0.3ex\hbox{$<$}\kern-0.75em{\lower0.65ex\hbox{$\sim$}}} 
\def\gsim{\raise0.3ex\hbox{$>$}\kern-0.75em{\lower0.65ex\hbox{$\sim$}}} 
\def\sc1{\raise2.1ex\hbox{\tiny $r\!\!=\!\!4$}\kern-0.95em{\hbox{$=$}}}

\def\cm3{~{\rm cm^{-3}}}

\def\ltsima{$\; \buildrel < \over \sim \;$}
\def\simlt{\lower.5ex\hbox{\ltsima}}
\def\gtsima{$\; \buildrel > \over \sim \;$}
\def\simgt{\lower.5ex\hbox{\gtsima}}

\def\sc{{\it Science\ }}

\def\ion#1#2{#1$\;${\small\rm\@Roman{#2}}\relax}

\newcount\lecurrentfam
\def\LaTeX{\lecurrentfam=\the\fam \leavevmode L\raise.42ex
\hbox{$\fam\lecurrentfam\scriptstyle\kern-.3em A$}\kern-.15em\TeX}
\def\sizrpt{
(\fontname\the\font): em=\the\fontdimen6\font, ex=\the\fontdimen5\font
\typeout{
(\fontname\the\font): em=\the\fontdimen6\font, ex=\the\fontdimen5\font
}}
\def\refindent{\advance\leftskip by 24pt \parindent=-24pt}
\def\appgdef#1#2{%
 \@ifxundefined#1{\toks@{}}{\toks@\expandafter{#1}}%
 \toks@ii{#2}%
 \xdef#1{\the\toks@\the\toks@ii}%
}%
\def\@tablenotetext#1#2{%
 \vspace{.5ex}%
 {\noindent\llap{$^{#1}$}#2\par}%
}%
\def\@tablenotes#1{%
 \par
 \vspace{4.5ex}\footnoterule\vspace{.5ex}%
 {\footnotesize#1}%
}%
\def\@tablenotes@pptt#1{%
 \par
 \vspace{3.2ex}\footnoterule\vspace{.4ex}%
 {\footnotesize#1}%
}%
\AtBeginDocument{%
}%
\newcommand\tablenotetext[2]{%
 \appgdef\tblnote@list{\@tablenotetext{#1}{#2}}%
}%
\def\spew@tblnotes{%
 \@ifx@empty\tblnote@list{}{%
  \@tablenotes{\tblnote@list}%
  \global\let\tblnote@list\@empty
 }%
}%